\documentclass[
prc,preprint,nofootinbib,preprintnumbers,
superscriptaddress
]{revtex4-1}

\usepackage[usenames,dvipsnames,svgnames,table,x11names]{xcolor}
\usepackage[latin2]{inputenc} % for names in bibliography

\usepackage[colorlinks,pdfstartview=FitH,bookmarksnumbered,bookmarksopen]{hyperref}
\hypersetup{linkcolor=blue,citecolor=blue,filecolor=black,urlcolor=blue}

\usepackage{custom}

\graphicspath{{figures/}}

\usepackage[margin=1in,bottom=1in,footskip=.7in]{geometry}

\begin{document}

\title{
Hydro+: hydrodynamics with parametric slowing down
and fluctuations near the critical point
}

\author{M.~Stephanov}
\affiliation{Physics Department, University of Illinois at Chicago,
  W. Taylor St., Chicago IL 60607-7059}
\author{Y.~Yin}
\affiliation{Center for Theoretical Physics, Massachusetts Institute of Technology, Cambridge, MA 02139}

\date{\today}
\preprint{MIT-CTP/4969}
\begin{abstract}
  The search for the QCD critical point in heavy-ion collision
  experiments requires dynamical simulations of the bulk evolution of
  QCD matter as well as of fluctuations.   We consider two
  essential ingredients of such a simulation: a generic extension of
  hydrodynamics by a {\em parametrically\/} slow mode or modes
  (``Hydro+'') and a description of fluctuations out of equilibrium.
   By combining the two ingredients we are able to
  describe the bulk evolution and the fluctutations within the same
  framework.  Critical slowing down means that equilibration of
  fluctuations could be as slow as hydrodynamic evolution and thus
  fluctuations could significantly deviate from equilibrium near the
  critical point.  We generalize hydrodynamics to
  partial-equilibrium conditions where the state of the system is
  characterized by the off-equilibrium magnitude of fluctuations in
  addition to the usual hydrodynamic variables -- conserved
  densities. We find that the key  element  of the new
  formalism -- the extended entropy taking into account the
  off-equilibrium fluctuations -- is remarkably similar to the 2PI
  action in quantum field theory. We show how the new Hydro+ formalism
  reproduces two major effects of critical fluctuations on the bulk
  evolution: the strong frequency dependence of the anomalously large
  bulk viscosity as well as the stiffening of the equation of state
  with increasing frequency or wave-number. While the agreement with
  known results confirms its validity, the fact that Hydro+ achieves
  this within a local and deterministic framework gives it significant
  advantages for dynamical simulations.
\end{abstract}

\maketitle

\section{Introduction and outline}

Hydrodynamics is a universal and versatile theory which describes (as
the name implies) the dynamics of fluids or, more generally, systems
whose evolution at long distance and time scales essentially consists
of redistribution of conserved quantities (energy, momentum and
charges) towards achieving global thermodynamic equilibrium throughout
the system. Although many results of this paper are general, its
primary focus is the application of relativistic hydrodynamics to the
evolution of the fireball of hot and dense QCD matter created in
heavy-ion collisions. In particular, this work is motivated by one of
the major goals of the heavy-ion collision experiments -- the discovery of
the QCD critical point through the beam energy scan of the QCD phase
diagram \cite{Stephanov:1998dy,Stephanov:2004wx}.  Hydrodynamic
description has shown remarkable and non-trivial agreement with many
results of heavy-ion collision experiments (see,
e.g.,~\cite{Heinz:2013wva} for a concise review and further
references). So far, however, such applications of hydrodynamics have
been limited to the regime where the equation of state does not
contain a critical point, e.g., at negligible baryon number
densities. In order to facilitate the search for the QCD critical
point it is essential to extend hydrodynamic description of heavy-ion
collisions into the regime of finite baryon densities and,
specifically, into the vicinity of the critical point.

There is, however, a major problem with applying hydrodynamics near a
critical point. 
Applicability of hydrodynamics rests on the possibility of a certain scale separation. The time scales for
achieving {\em local} equilibrium are usually much shorter than the time
scales needed to reach global equilibrium throughout the system. 
This scale separation exists because the conserved densities relax to equilibrium by
diffusion and the relaxation time is proportional to the square of the
size of the inhomogeneity involved. 
The local equilibrium is achieved at time scales necessary to smooth out inhomogeneities on the scale of
the correlation length $\xi$ which is, typically, microscopically
small. The global equilibrium may require times which are arbitrarily
long for arbitrarily large systems. This separation of scales
disappears at the critical point as the correlation length $\xi$
diverges (see, e.g., Ref.~\cite{RevModPhys.50.85} for more discussion).

At first sight, it might seem that hydrodynamics works as long as the
correlation length~$\xi$, while becoming large, remains much shorter
than the scale of the inhomogeneities, $\ell$. For heavy-ion
collisions where the relevant size of the system is of order 10 fm
while the correlation length is most likely not to exceed
$2-3$~fm~\cite{Stephanov:1999zu,Berdnikov:1999ph} one would then
expect a reasonable scale separation. However, such an argument would
miss an important point: the time for relaxation of inhomogeneities of
size $\xi$, essential for establishing local thermodynamic
equilibrium, grows faster than their size and become very long near
the critical point (critical slowing down).  This time grows with
$\xi$ as $\xi^z$, where $z\approx 3$~\cite{RevModPhys.49.435}.  Since
$z>1$, hydrodynamics breaks down even earlier, at much longer time and
distance scales than the naive argument would suggest. For example, a
sound wave with period shorter than the time of relaxation to local
equilibrium cannot be described hydrodynamically, which happens when
the wavelength $\ell\sim\xi^3$ (in units of temperature) or shorter, which is
much larger than just $\xi$.  For heavy-ion collisions the time scales
needed to reach local equilibrium could become comparable to typical
evolution times, invalidating hydrodynamic description near the
critical point well before $\xi$ is as large as $\ell$.

One way this breakdown is manifested is in the growth of gradient
corrections to the constitutive equation for the stress tensor, i.e.,
non-equilibrium corrections to pressure, proportional to bulk
viscosity $\zeta$ and  expansion rate (gradient of velocity) 
$\theta=\bm{\nabla\cdot v}\sim 1/\ell$.
%: $-\zeta \bm{\nabla\cdot v}$.
The bulk viscosity diverges at the critical point as
$\zeta\sim \x^{z-\alpha/\nu}\sim \xi^3$ (for simplicity, we round
exponents to integer values). This leads to the breakdown of the
gradient expansion, i.e., of the locality of hydrodynamic description,
when $\zeta\theta \sim \xi^3/\ell\gtrsim 1$, i.e., already for $\ell \lesssim \xi^3$, i.e., earlier, at much larger $\ell$ than just
$\xi$ suggested by the naive argument.

The goal of this paper is to develop an extended hydrodynamic
description which overcomes this shortcoming of ordinary
hydrodynamics. One way to understand the predicament here is to
compare this situation with the breakdown of an effective theory for
low-energy modes at scales comparable to the energy scales at which
the mode which is next-to-lowest can get excited. The solution in this
case is well-known -- include the latter mode into the effective
theory description.  Therefore it is logical to consider extending
hydrodynamics by adding an additional mode or, rather, as we shall see,
a set of modes, describing relaxation processes  responsible
for the critical slowing down.

Of course, this only makes
sense if the mode we add is still {\em parametrically\/} slower than the
remaining (infinitely many) microscopic modes which are not included
explicitly.\footnote{Hydrodynamics is a truncation of the full theory,
  justified by the slowness of hydrodynamic degrees of freedom. Due to
  conservation, the relaxation rates of these modes are controlled by the
  scale of inhomogeneity $\ell$, and are proportional to
  $\ell^{-2}$. The rate for the slow mode we add should be also
  controlled by some parameter independently of $\ell$. } We shall consider the role of such
parametric separation in general and in the vicinity of the critical
point in particular, where it is controlled by the divergence of the
correlation length.

\subsection{Outline}

Broadly speaking, our paper contains two major  ingredients. They
are independent but, obviously, related.  The first is presented
in Section~\ref{sec:hydro-slow} and the second in
Sections~\ref{sec:entr-fluct-2pi}
and~\ref{sec:fluctuation-modes}. They are combined in
Section~\ref{sec:hydro+-near-qcd}.

First, in Section~\ref{sec:hydro-slow}, we consider a generic
extension of the relativistic hydrodynamics by a mode which is slow
not because it is a conserved density, but because a parameter
controlling its relaxation rate can be independently tuned to make the
rate arbitrarily small.  Such a model could describe
many different systems where due to {\em microscopic} dynamics some
relaxation processes are slow. For example, if an additional charge
exists which is only approximately conserved, so that its relaxation
is not diffusive, but nevertheless slow. In
Appendix~\ref{sec:nA} we discuss a specific example with partially
conserved axial charge playing this role.   Another example could be a
system where some channels of chemical equilibration (of relative
particle abundances) are slow. Such situations arise in cosmological
and astrophysical contexts due to slowness of electroweak
processes.

We use this simple model, which we shall call ``Hydro+'' to illustrate
how the competition between the two relaxation scales (hydrodynamic
and non-hydrodynamic) produces two regimes, one where ordinary
hydrodynamics is sufficient and another where Hydro+ is necessary. This
allows us to illustrate the basic mechanisms and in particular show
how the divergent bulk viscosity and corresponding breakdown of
the hydrodynamic gradient expansion is related to the slow mode. 
This mechanism is known \cite{mandelshtam1937theory,landau1954anomalous,landau2013fluid}, and our purpose here is to review and
present it in {\em relativistically covariant} form, setting the stage for
the generalization necessary near the QCD critical point.

To introduce the second ingredient we need to address the question of
what is the physical origin of the critically slow  processes  near the
critical point. It is known that  the critical slowing down affects
relaxation of fluctuations at scales of order $\xi$, essential for
establishing local equilibrium.  In order to
find an appropriate description of these processes, for our
purposes, we develop a more general approach to evolution of
fluctuations in hydrodynamics in Sections~\ref{sec:entr-fluct-2pi} and
~\ref{sec:fluctuation-modes}. 

In Section~\ref{sec:entr-fluct-2pi} we introduce the
notion of the partial-equilibrium entropy $S_2$ which is a functional not
only of the {\em average values} of the conserved densities, but also
of the {\em magnitude of their fluctuations} (and correlations), i.e.,
one- and two-point functions of these densities. We show that this
entropy has a remarkable (mostly mathematical) similarity to the  2-particle
irreducible (2PI) action in quantum field theory. 

Having derived the ``2PI entropy'' $S_2$ we then use it in
Section~\ref{sec:fluctuation-modes} to write evolution/relaxation
equations for the conserved densities (which are ordinary hydrodynamic
equations) together with the relaxation equations for the 2-point
correlators.  The equations for 2-point correlators show a remarkable
similarity to kinetic (Boltzmann) equations, as has been observed long
time ago in a related work by Andreev {\it et
  al.}~\cite{andreev1970twoliquid,andreev1978corrections,lebedev1981fluctuation,lebedev1982fluctuation},
or more recently, in the context of heavy-ion collisions  away from
the critical point,  in
Ref.\cite{Akamatsu:2016llw,Akamatsu:2017rdu}. Kinetic equations for correlations
functions have also appeared in the work of
Kawasaki~\cite{KAWASAKI19701} and others~\cite{RONIS198125,
  MACHTA1982361,SCHOFIELD199289} in the calculations of kinetic
coefficients and higher-order correlation functions near critical
points in non-relativistic systems. In this paper we use  a similar
 ``kinetic'' approach together with the 2PI entropy formalism we
introduced to extend applicability of hydrodynamics for simulations
near the QCD the critical
point.

To this end, we defer further development of the 2PI entropy formalism
to future work, and focus on its application to Hydro+ near the
critical point in Section~\ref{sec:hydro+-near-qcd}. We identify the
slowest mode of fluctuations and propose the equation which describes
its evolution as an extension/generalization of ordinary
hydrodynamics. To demonstrate how this extension works near the
critical point and to verify its validity we compare the
frequency-dependent bulk response in Hydro+ to the existing result due
to Onuki~\cite{PhysRevE.55.403, onuki2002phase} obtained using a
different approach based on a stochastic hydrodynamic model of
critical fluctuations. We find that the new formalism simplifies this
analytical calculation, while also providing a different intuitive
perspective. The main advantage of the new formalism is that it is
{\em local} and {\em deterministic}, while still capturing the dynamics of
fluctuations. These properties make it much easier to use for
simulating dynamics  of the heavy-ion collision fireball expanding in the
vicinity of the QCD critical point.

\section{Hydrodynamics with an additional slow mode
\label{sec:hydro-slow}
}

\subsection{The general framework}

In this section, we will formulate a local description of the
evolution of hydrodynamical variables such as energy density $\ed$, a
conserved charge (e.g., baryon number) density~$n$ and fluid velocity
$u^{\mu}$ coupled to \textit{one} additional non-hydrodynamic, but
nevertheless slow, scalar mode which will be denoted by~$\phi$  (see
also Ref.~\cite{Stephanov:2017wlw}).

If the mode $\phi$ is {\em parametrically} slower than other
microscopic modes we can, in a window of time scales, consider
{\em partial-equilibrium} state in which the entropy reaches its maximum
under an additional constraint that the expectation value of that slow
mode is $\phi$.  We can then introduce partial-equilibrium entropy
$s_{(+)}(\ed, n , \phi)$ as the logarithm of the number of states
satisfying this constraint.  The generalized thermodynamic potentials
are defined as usual via derivatives of entropy
\begin{equation}
\label{ds++}
d s_{(+)}
=\b_{(+)}\, d\ed -\a_{ (+)}\, dn - \pi d\,\phi\, .
\end{equation}
Here $\b_{(+)}(\ed,n,\phi)$ and $\a_{(+)}(\ed,n,\phi)$ are generalized
inverse temperature and chemical potential to temperature ratio,
respectively, in the partial-equilibrium state. % The quantities $\b_{(+)}$
% and $\a_{(+)}$ are functions of $\ed, n, \phi$. 
The variable
$\pi(\ed, n, \phi)$ is the generalized thermodynamic potential (or
``force'') corresponding to $\phi$. In a {\em complete} equilibrium at given
$\ed$ and $n$ the variable $\phi$ must relax to its equilibrium value
$\bar\phi(\ed,n)$ which maximizes the generalized entropy, i.e.,
\begin{equation}
  \label{eq:pi=0}
  s(\ed,n) = \max_\phi s_\plus(\ed,n,\phi) = s_\plus (\ed,n,\bar\phi(\ed,n)),
\end{equation}
and thus
\begin{equation}\label{eq:pibar}
\pi(\ed,n,\bar\phi(\ed,n)) =0.
\end{equation}
Hydro+ equations are usual energy-momentum conservation
$\pd_{\mu}\, T^{\mu\nu}=0$, charge conservation $\pd_{\mu}\,
J^{\mu}=0$ and an additional equation which describes the relaxation
of $\phi$ towards equilibrium that we will specify shortly.

The components  of $T^{\mu\nu}$ must be local functionals of the variables
$u^\mu$, $\ed$, $n$, $\phi$. 
One can expand in powers of derivatives, as usual:
\begin{equation}
\label{Tmunu}
  T^{\mu\nu} =  \ed u^\mu u^\nu + p_{(+)}\, g_\perp^{\mu\nu} + \DT^{\mu\nu}\, ,
\end{equation}
where
\begin{equation}
 \label{eq:g-perp}
  g_\perp^{\mu\nu} = g^{\mu\nu} +u^\mu u^\nu 
\end{equation}
is the transverse (spatial in the local rest frame) part of
$g^{\mu\nu}$ \footnote{We use the mostly positive convention for the
  metric, therefore, $u\cdot u=-1$.} and the function
$p_{(+)}(\ed,n,\phi)$ is the generalized partial-equilibrium
pressure. %expressed as a function of $\ed$, $n$ and $\phi$.
Here $\D T^{\mu\nu}$ denotes contributions to the stress-energy tensor
due to the gradients of $u^\mu$, $\ed$, $n$, $\phi$,  which vanish
in a static homogeneous system.  Throughout this paper, we will use
Landau frame choice to define $\ed$ and 4-velocity~$u^\mu$, i.e.,~$u_{\mu}\, \D T^{\mu\nu}=0$.
Similarly, the definition $n=u_{\mu}\, J^{\mu}$ implies:
\begin{equation}
J^{\mu} = n\, u^{\mu} + \D J^{\mu}\,  
\end{equation}
with $u\cdot\Delta J =0$.
Again $\D J^{\mu}$ will vanish in a static and homogeneous system and
can be expanded in powers of derivatives. 

The five equations for hydrodynamic variables can now be written explicitly:
\bes
\label{hydro+}
\begin{eqnarray}
\label{De}
D\e &=& - w_{(+)} \theta 
- \(\pd_{\mu}u_{\nu}\)\DT^{\mu\nu}_{\perp}\, ,
\label{Dn}
\\
Dn&=& -n\,\theta\,
-\pd\cdot \D J\, , 
\\
\label{Du}
w_{(+)} D u^{\nu}&=&-\pd^{\nu}_{\perp} p 
- {\delta_{\perp}}^\nu_\lambda\pd_{\mu}\, \Delta T^{\mu\lambda}\, .
\end{eqnarray}
The equation for $\phi$ must describe relaxation of this quantity to
equilibrium value $\bar\phi(\ed,n)$ and can be written as
\begin{equation}
  \label{phi-eq}
D\,\phi
  = - F_{\phi} -A_{\phi}\, \theta\,  ,
\end{equation}
\ees Here $F_{\phi}(\ed,n,\phi)$ is the ``returning force'' which, at given $\ed$
and $n$, drives $\phi$ back to its equilibrium value. I.e., $F_\phi=0$
when $\pi=0$ (see Eq.~(\ref{eq:pi=0})).  The coefficient
$A_{\phi}(\ed,n,\phi)$
describes the susceptibility of the quantity $\phi$ to isotropic
compression/expansion.    In \eqref{hydro+}, we also introduced
notations
\begin{equation}\label{eq:Dtheta}
w_{(+)}\equiv\ed+p_{(+)},\quad D\equiv u\cdot\pd \quad\mbox{ and }\quad \theta \equiv\pd\cdot u\, .
\end{equation}

For conventional hydrodynamics, the system of equations is closed once
we supply the equation of state and constitutive relations for $\D
J_{\mu}$ and $ \Delta T^{\mu\nu}$.  Similarly, Hydro+ will be
closed if we, in addition,
supply %the generalized pressure $p_\plus(\ed, n, \phi)$ and
$F_{\phi}(\ed,n,\phi)$ and $A_\phi(\ed,n,\phi)$.

Second law of thermodynamics imposes constraints on the form of
the constituitive equations. For ordinary hydrodynamics it requires
$\beta p = s
-\beta\ed +\alpha n$ and positivity of kinetic coefficients in $\Delta
J^\mu$ and $\Delta T^{\mu\nu}$. Similarly, there are constraints on
$p_{(+)}(\ed, n, \phi)$, $F_{\phi}(\ed,n, \phi)$ and
$A_\phi(\ed,n,\phi)$ from the generalized second law of thermodynamics
which requires $\pd^{\mu}\, s^{\plus}_{\mu}\geq 0$.  The generalized (partial-equilibrium)
entropy current is given by
\begin{equation}
  \label{eq:s-mu}
  s^\mu_{\plus} = s_{\plus}\, u^\mu 
+  \D s^{\mu} \, ,
\end{equation}
where $\D s^{\mu}$ denotes contribution to entropy current from gradients.  
The divergence of the entropy current then becomes:
\begin{multline}
  \label{eq:del-s}
  \del_\mu s_\plus^\mu = 
  \(s_{\plus}-\b_{\plus}\,w_{\plus}+\a_{\plus}\, n+\pi A_{\phi}\)\, \theta
   \\
-
\a_{\plus} \(\pd \cdot \D J \)
+\pi \, F_{\phi} + \b_{\plus}\(\pd_{\mu}u_{\nu}\)\, \D T^{\mu\nu}_{\perp} +\pd_{\mu}\, \(\D s^{\mu}+\a_{\plus}  \D J^\mu \)\, ,
\end{multline}
where we used \eqref{ds++} and Eq.~(\ref{Du}), while neglecting
terms cubic in gradients.  In order to guarantee $\pd_{\mu}\,
s_\plus^{\mu}\geq 0$, we need the first term on the R.H.S of
\eqref{eq:del-s} to vanish.  This condition relates pressure
$p_{\plus}$ and ``compressibility'' $A_{\phi}$ and other generalized
thermodynamic functions such as $s_{\plus}$, $\a_{\plus}$, $\b_{\plus}$:
\begin{equation}
\label{p+}
\b_{\plus} p_{\plus}= \, s_{\plus}-\b_{\plus} \ed  +\a_{\plus}\, n +\pi\, A_{\phi}\, . 
%\qquad
%p \equiv w -\e\, . 
\end{equation}
Therefore:
\begin{equation}
\label{beta-dp}
\b_{\plus}\, d p_{\plus} = -w_{\plus}\, d\b_{\plus} + n\, d\a_{\plus} - \pi\, d\phi + d\(\pi A_{\phi}\)\, , 
\end{equation}
where we used \eqref{ds++}. 
The last term in Eq.~(\ref{eq:del-s}) must also vanish, which
requires $\Delta s^\mu = -\a_\plus\D J^\mu$, similarly to ordinary hydrodynamics.

The dissipative terms
$\D T^{\mu\nu}$, $\D J^{\mu}$ and $F_{\phi}$ are also constrained
by the second law of thermodynamics similarly to ordinary hydrodynamics. 
% The mixing between scalar mode $\a, \b,\phi$ and $\D
% T^{\mu\nu}_{\perp}$ begins at the second order in gradients. 

To the first order in gradients, we still find the usual form for the
gradient corrections to stress energy tensor:
\begin{equation}
\label{DT}
\D T^{\mu\nu} 
= -\eta_{\plus} \(\pd_\perp^\mu u^{\nu} + \pd_\perp^\nu u^{\mu} 
- \frac23 g^{\mu\nu}_{\perp} \theta\) 
-\zeta_{\plus} g^{\mu\nu}_{\perp}\, \theta\, . 
\end{equation}
The second law of thermodynamics requires that $\zeta_{\plus},
\eta_{\plus}\ge0$.

To first order in gradients, we now have additional term in $\Delta J^\mu$
\begin{eqnarray}
\label{DJ+}
\D J^{\mu} %- \frac{n}{w_{\plus}}\, q^{\mu}
= - \l_{\a\a}\, \pd^{\mu}_{\perp}\, \a - \l_{\a\pi}\, \pd^{\mu}_{\perp}\, \pi\,  
\end{eqnarray}
and
\begin{equation}
  \label{eq:Fphi}
F_\phi = \gamma_{\pi} \pi - \partial_\perp \cdot \(\, \lambda_{\pi\pi}\,\partial\pi+  \lambda_{\alpha\pi}\, \partial\alpha_{\plus}\,\)\, ,
\end{equation}
with $\g_{\pi}\ge0$ and a semi-positive definite matrix $\l_{ab}$
($a,b=\a,\pi$).  Eqs.~\eqref{DJ+} and \eqref{eq:Fphi}, take into
account Onsager reciprocity.  The first
term in $F_\phi$ is allowed because $D\phi$ can remain finite even if
the system is homogeneous since $\phi$ is {\em not} a conserved
quantity.

Equation~(\ref{phi-eq}) for $\phi$ reads, upon substituting Eq.~(\ref{eq:Fphi}):
\begin{eqnarray}
\label{Dphi1}
D\phi = - \g_{\pi}\, \pi- \partial_\perp \cdot \(\, \lambda_{\pi\pi}\,\partial\pi+  \lambda_{\alpha\pi}\, \partial\alpha_{\plus}\,\)\, . 
\end{eqnarray}

At sufficiently long times, i.e., in the limit $\pi=0$, Eq.~\eqref{DJ+}
reproduces the conventional constitutive relation for the dissipative
part of $J^\mu$ with $\l_{\a\a}$ giving the conventional
conductivity (times temperature), $\sigma T$, i.e.
$\Delta J^{\mu}=-\sigma T \pd^{\mu}_{\perp}\, \a$.  In
Appendix~\ref{sec:nA} we use hydrodynamics with partially conserved
axial charge as an example of a theory with nonzero $\g_{\phi}$,
$\l_{\a\a}$, $\l_{\a\pi}$, and $\l_{\pi\pi} $.

It is instructive to compare and contrast the single-mode Hydro+ and
the model of chiral fluid dynamics (CFD) (see, e.g.,
\cite{Nahrgang:2011mg,Herold:2013bi}) considered recently in the
context of the QCD critical point. The equation (linearized for
simplicity) of the non-hydrodynamic mode $\sigma$,
$\partial\cdot\partial \sigma + \eta_\sigma \partial_t \sigma +
m_\sigma^2\sigma=0$, is different from the (correspondingly
linearized) equation~(\ref{Dphi1}). Most notably, the mode $\sigma$ is
propagating, not relaxational, unless, or course, one considers
$\omega\ll\eta_\sigma$. In this case the relaxation rate
$\Gamma_\pi=m_\sigma^2/\eta_\sigma$ is vanishing when $m_\sigma\to0$,
provided $\eta_\sigma$ does not vanish in this limit, which, however,
it does in the model. If one ignores the physics of the model and
considers $\eta_\sigma$ as a free phenomenological parameter then the
model will become an example of a single-mode Hydro+, provided the
Lorentz invariance is also restored by replacing
$\partial_t\sigma\to D\sigma$. This should be expected since Hydro+ is
a general effective theory which should match any model in the
appropriate limit, the specifics of the model being reflected in the
values of phenomenological parameters such as $\gamma_\pi$, $A_\phi$,
etc.   It should be also noted that, due to the mixing of the scalar
field with baryon density, addition of such a field to hydrodynamics,
generally, will not produce an independently slow mode even when $m_\sigma\to0$
\cite{Son:2004iv} unless an additional parameter is tuned. 

Summarizing this section, we have considered a generalization of
hydrodynamics, or ``Hydro+'', which describes the coupled evolution of
hydrodynamic degrees of freedom and an additional parametrically slow
scalar mode.  As in ordinary hydrodynamics, the second law of
thermodynamics imposes constraints on the form and parameters of the
theory.  The inputs of Hydro+ include the generalized entropy
$s_{\plus}(\ed,n,\phi)$, the ``compressibility coefficient''
$A_{\phi}$ and transport coefficients such as
$\eta_{\plus}, \zeta_{\plus}$ and
$\g_{\pi}, \l_{\a\a},\l_{\a\pi}, \l_{\pi\pi} $ which appear in
constitutive equations \eqref{DT},~(\ref{DJ+}) and~\eqref{eq:Fphi}.

For very slow processes, i.e., processes slower than the relaxation
time of the slow mode (equivalently, at $\pi=0$), Hydro+ reduces to
conventional hydrodynamics. While the relationship between
$\eta_\plus$ and $\lambda_{\alpha\alpha}$ to the conventional
hydrodynamic coefficients $\eta$ and $\lambda$ (given by Kubo formulas
at $\omega\to0$) is trivial, this is not the case for the bulk
viscosity $\zeta$ because it receives contribution from the slow mode
proportional to its large relaxation time, as pointed out long ago by
Leontovich and
Mandelstam~\cite{mandelshtam1937theory,landau1954anomalous,landau2013fluid}.
%,%$\zeta =\zeta_\plus + \D\zeta$, 
In the next section we discuss
this effect in more detail using Hydro+ with a single slow non-hydrodynamic
mode. Generalization of this effect to the case of the critical point
leads to the critical divergence of the bulk viscosity which we
discussed already in the introduction.

\subsection{Bulk viscosity and sound in Hydro+}
\label{sec:plus-dispersion}

The presence of the slow mode has a profound effect on
the response of the system to expansion or compression. In
hydrodynamics the fluid's expansion/compression leads to the
corresponding change in the densities of the conserved quantities. If
the size of the system (or its part) undergoing expansion/compression
is large enough that the contribution of diffusive processes are
negligible, the conserved quantities, energy and charge, remain the
same and the densities, $\ed$ and $n$, simply scale with the
volume. We can describe this process by the linearized
Eqs.~(\ref{De}),~(\ref{Dn}), where $\theta=\bm{\nabla\cdot v}$ is the
expansion rate:
\begin{equation}
  \label{eq:en-linear}
  D \varepsilon = -w\theta + \ldots
\qquad\mbox{and}\quad
D n = -n\theta + \ldots
\end{equation}
where the ellipsis denotes the terms of higher order in gradients.
% The entropy is also conserved, and thus the entropy density obeys
% \begin{equation}
%   \label{eq:s-linear}
%   D s = -s\theta + \ldots
% \end{equation}

The pressure, on the other hand, is a function of the variables $\ed$
and $n$ only in equilibrium (at $\pi=0$). In ordinary hydrodynamics
the pressure adjusts to its equilibrium value $p(\ed,n)$ on a
microscopically short time scale, negligible compared to the timescale
of expansion. In Hydro+, in contrast, the pressure $p_\plus$ depends
also on the variable $\phi$ or, $\pi$, whose relaxation rate to
equilibrium, $\pi=0$, can be arbitrarily slow. As a result, if we write
the linearized deviation of pressure from equilibrium due to
infinitesimal expansion/compression $\theta$ we find an additional
term proportional to $\pi$:
\begin{equation}
  \label{eq:p-pi}
  p_\plus (\ed,n,\pi) = p(\ed,n) + p_\pi(\ed,n) \pi + \ldots
\end{equation}
The deviation of $\pi$ from equilibrium is due to expansion, and the
amount of this deviation is proportional to $\theta$. To express this
explicitly, we can substitute $\phi(\ed,n,\pi)$ into
Eq.~(\ref{phi-eq}) to rewrite it as an equation for $\pi$:
\begin{equation}
  \label{eq:pi-eq}
  \phi_\pi D\pi = -\gamma_\pi \pi 
+ \left[
 w \td{\phi}{\ed}_{n\pi} +  n \td{\phi}{n}_{\ed\pi} - A_\phi 
\right]\theta  + \ldots
\end{equation}
where we used Eqs.~(\ref{eq:en-linear}) and defined
\begin{eqnarray}
\label{phi-pi-def}
 \phi_\pi\equiv \(\frac{\pd \phi}{\pd \pi}\)_{\ed n}\, . 
\end{eqnarray}

Using Maxwell relations (see Appendix~\ref{sec:therm-relat-p_pi})
one can express the quantity in the square brackets in terms of $p_\pi$:
\begin{equation}
  \label{eq:p-pi-phi-ed}
 \beta p_\pi = - \left[
 w \td{\phi}{\ed}_{n\pi} +  n \td{\phi}{n}_{\ed\pi} - A_\phi
\right]
\end{equation}
and rewrite Eq.~(\ref{eq:pi-eq}) as
\begin{equation}
  \label{eq:Dpi-Gamma-ppi}
  D\pi = -\Gamma_\pi \pi - \frac{\beta p_\pi}{\phi_\pi}\theta 
+ \ldots
\end{equation}
where we defined the relaxation rate
\begin{eqnarray}
\label{G-phi-def}
\G_\pi &\equiv& \frac{\g_{\pi}}{\phi_\pi}\,.
\end{eqnarray}

This linearized equation can be solved for $\pi$ as
\begin{equation}
  \label{eq:pi-theta}
  \pi = \frac{\beta p_\pi}{\phi_\pi}\frac{1}{i\omega-\Gamma_\pi}\theta
\end{equation}
where $\omega$ is the frequency of the oscillation of the variables
around equilibrium. Substituting into Eq.~(\ref{eq:p-pi}) we find that
pressure deviates from its equilibrium value by the amount
proportional to the expansion rate $\theta$:
\begin{equation}
  \label{eq:p_plus-theta}
  p_\plus = p 
- \frac{\beta p_\pi^2}{\phi_\pi}\frac{1}{\Gamma_\pi - i\omega}\theta\,.
\end{equation}
By definition, the coefficient of $\theta$ at $\omega=0$ is the
contribution of the slow mode to the bulk viscosity:
\begin{equation}
  \label{eq:zeta0}
  \Delta\zeta(0) = \frac{\beta p_\pi^2}{\phi_\pi\Gamma_\pi}.
\end{equation}
This contribution diverges when $\Gamma_\pi$ vanishes.

\begin{figure}
\subfigure[]
{
\includegraphics[width=.45\textwidth]{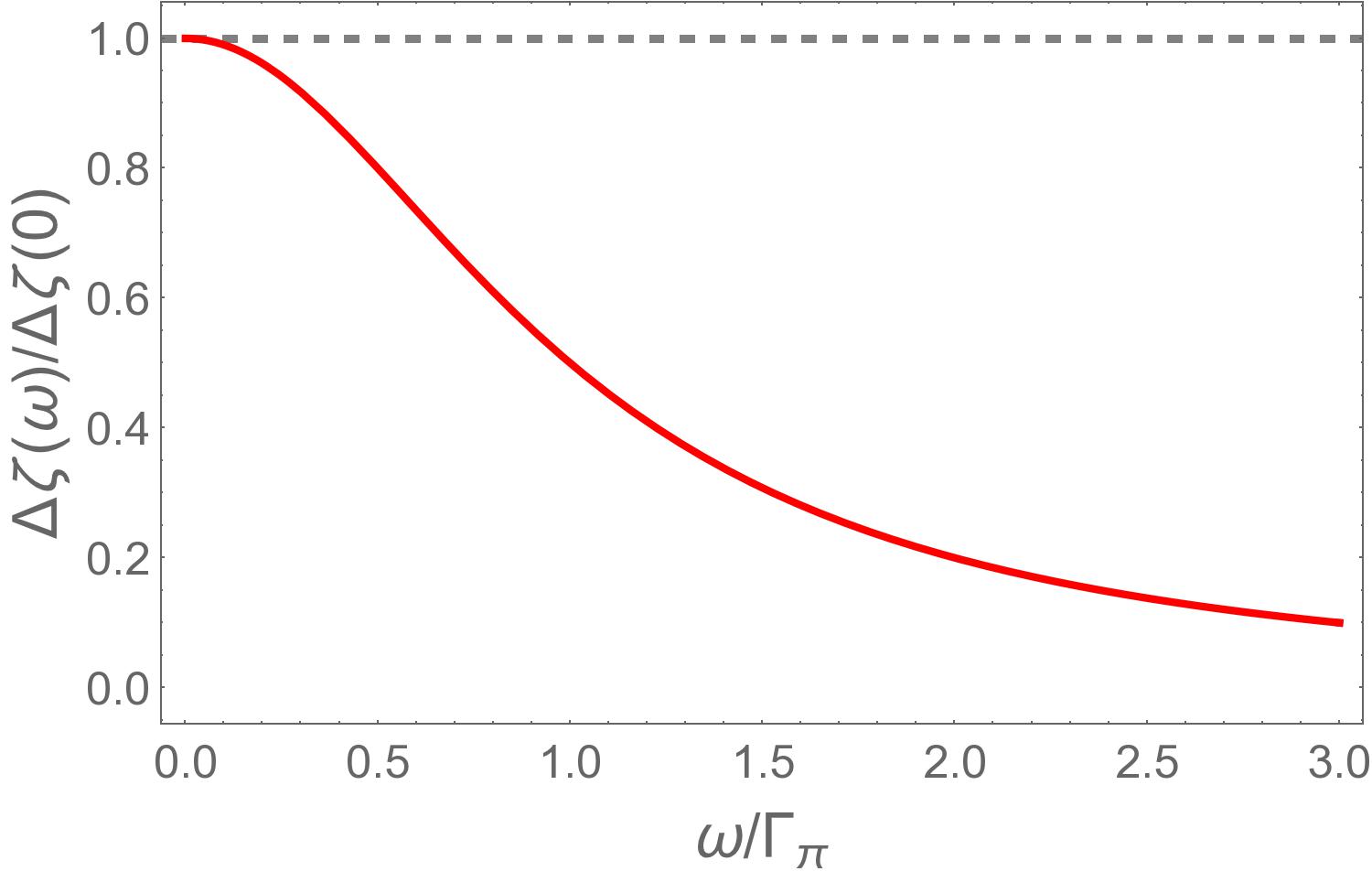}
\label{fig:ImG}
}
\subfigure[]
{
\includegraphics[width=.45\textwidth]{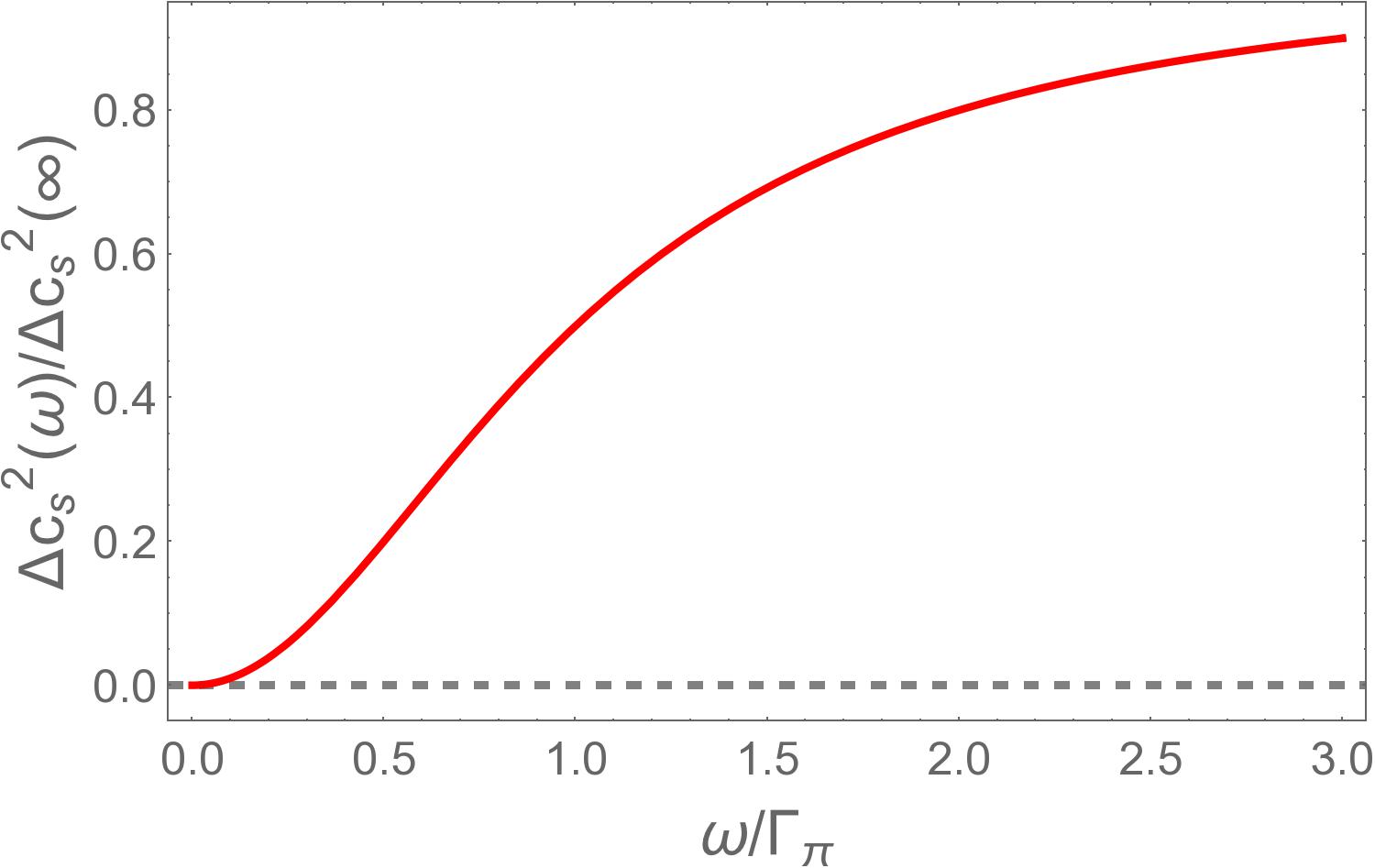}
\label{fig:ReG}
}
\caption[]{\label{fig:GR1mode}
 (color online)  The frequency dependence of the contribution of the
 slow mode to the bulk viscosity, or to
 $-{\rm Im}\,G_R/\omega$, (left), as well as to the speed of sound, or ${\rm
   Re}\,G_R/\o$, (right).
Dashed horizontal lines represent the (lack of) frequency dependence of this
quantities in ordinary hydrodynamics. 
  }
\end{figure}

At non-zero $\omega$ the coefficient of $\theta$ can be related to the
Green's function $G_R$ of the operator $T^i_i/3$, which is natural since it
describes the response to compression: 
\begin{equation}
  \label{eq:GR}
\frac{\beta p_\pi^2}{\phi_\pi}\frac{1}{\Gamma_\pi-i\omega}
  \equiv
\frac{i\Delta G_R(\omega)}{\omega}\,.
\end{equation}
The frequency dependent bulk viscosity can be defined as the real part
of that coefficient, or $-{\rm Im}\, G_R/\omega$, in accordance
with the Kubo formula,
\begin{equation}
  \label{eq:zetaomega}
  \Delta\zeta(\omega) = \frac{-{\rm Im}\,\Delta G_R(\omega)}{\omega} 
= \Delta\zeta(0)\,\frac{\Gamma_\pi^2}{\Gamma_\pi^2+\omega^2}\,.
\end{equation}
This quantity describes dissipation during 
expansion/compression at frequency $\omega$. Note that $\D\zeta(\omega)$
drops off when $\omega\gtrsim  \Gamma_\pi$. This means that if we were to naively extend the
conventional hydrodynamics with frequency-independent $\zeta$ to
$\omega\gtrsim \Gamma_\pi$ we would overestimate the amount of
dissipation (see
Fig.~\ref{fig:ImG}).

%%%%%%%%%%%%%%%%%%%%%%%%%%%%%%%%%
%% 
%%.   Plot GR 
%%
%%
%%%%%%%%%%%%%%%%%%%%%%%%%%%%%%%%%

The imaginary part of the coefficient of $\theta$ in
Eq.~(\ref{eq:p_plus-theta}), i.e., the real part of $\Delta G_R$ in
Eq.~(\ref{eq:GR}), is related to the contribution of the slow mode to
the sound speed. This can be seen from Eq.~(\ref{eq:p_plus-theta})
by expressing oscillation $\delta p$ of pressure in terms of
oscillations $\delta\ed$ of energy density and also using
Eq.~(\ref{eq:en-linear}) to express $\theta$ in terms of $\delta\ed$
($\theta=i\omega\,\delta\ed/w$):
\begin{equation}
  \label{eq:delta-p-plus}
  \delta p_\plus = \left(
c_s^2 + \Delta G_R/w 
%\D c_s^2(\omega)
%+ \frac{\beta p_\pi^2}{w\phi_\pi} 
%\frac{i\omega}{i\omega - \Gamma_\pi}
\right) \delta\ed\,.
\end{equation}
The expression $c_s^2+\D G_R/w$ in Eq.~(\ref{eq:delta-p-plus}) can be
viewed as the speed of sound (squared), or equation of
state stiffness, in two limits when it becomes almost real, with the imaginary
part related to the attenuation of the sound.

Let us define the frequency-dependent contribution to the speed of
sound (corresponding to phase velocity)
\begin{equation}
  \label{eq:Dcs}
  \Delta c_s^2(\omega) = \frac{{\rm Re}\,\Delta G_R(\omega)}{w}
= \frac{\beta p_\pi}{\phi_\pi w}\,\frac{\omega^2}{\Gamma^2+\omega^2}
\end{equation}
(see also Fig.~\ref{fig:ReG}).
The imaginary part (attenuation) becomes negligible in the limit $\omega\ll\Gamma_\pi$,
when the sound speed is given by $c_s^2$ -- the usual hydrodynamic
sound speed, and also in the limit $\omega\gg\Gamma_\pi$ when the
sound speed is given by a larger value
\begin{equation}
  \label{eq:cs_plus}
  {c_s^2}_\plus = c_s^2 + \Delta c_s^2(\infty), 
\qquad\mbox{where}\quad
\Delta c_s^2(\infty) = \frac{\beta p_\pi^2}{w\phi_\pi}\,.
\end{equation}
Comparing
Eq.~(\ref{eq:zeta0}) and~(\ref{eq:cs_plus}) we find:
\begin{equation}
\zeta(0) = w\,\frac{\Delta c_s^2(\infty)}{\Gamma_\pi}\label{eq:LK}
\end{equation}
-- the Landau-Khalatnikov formula (cf. Ref.\cite{landau2013fluid}).

Note that $\Delta c_s^2>0$ is a consequence of thermodynamic
stability. The fact that $c_\plus^2>c_s^2$, i.e., the Hydro+ equation
of state is stiffer, is natural since some (slow) degrees of freedom
are effectively ``frozen'' at high frequencies. Thus naively extending
ordinary hydrodynamics with equilibrium equation of state to higher
frequencies would underestimate the stiffness (see
Fig.~\ref{fig:ReG}).~\footnote{Also note that, while the hydrodynamic
  speed of sound $c_s^2$ is given by the usual derivative
  $(\partial p/\partial\ed)$ at $d\ed/w=dn/n$ and $\pi=0$, the speed
  ${c_s^2}_\plus$ is given by the derivative where instead of $\pi=0$
  the condition $ d\phi/A_\phi = d\ed/w = dn/n$ holds (see
  Appendix~\ref{sec:therm-relat-p_pi}). This is different from the
  $\phi={\rm const}$ condition in Ref.\cite{landau2013fluid} because
  even if the relaxation term in Eq.~(\ref{phi-eq}) can be neglected
  in the Hydro+ regime ($\omega\gg\Gamma_\pi$), the variable $\phi$
  oscillates with $\theta$ if its compressibility $A_\phi$ is
  nonzero.}

\begin{figure}
\subfigure[]
{
\includegraphics[width=.45\textwidth]{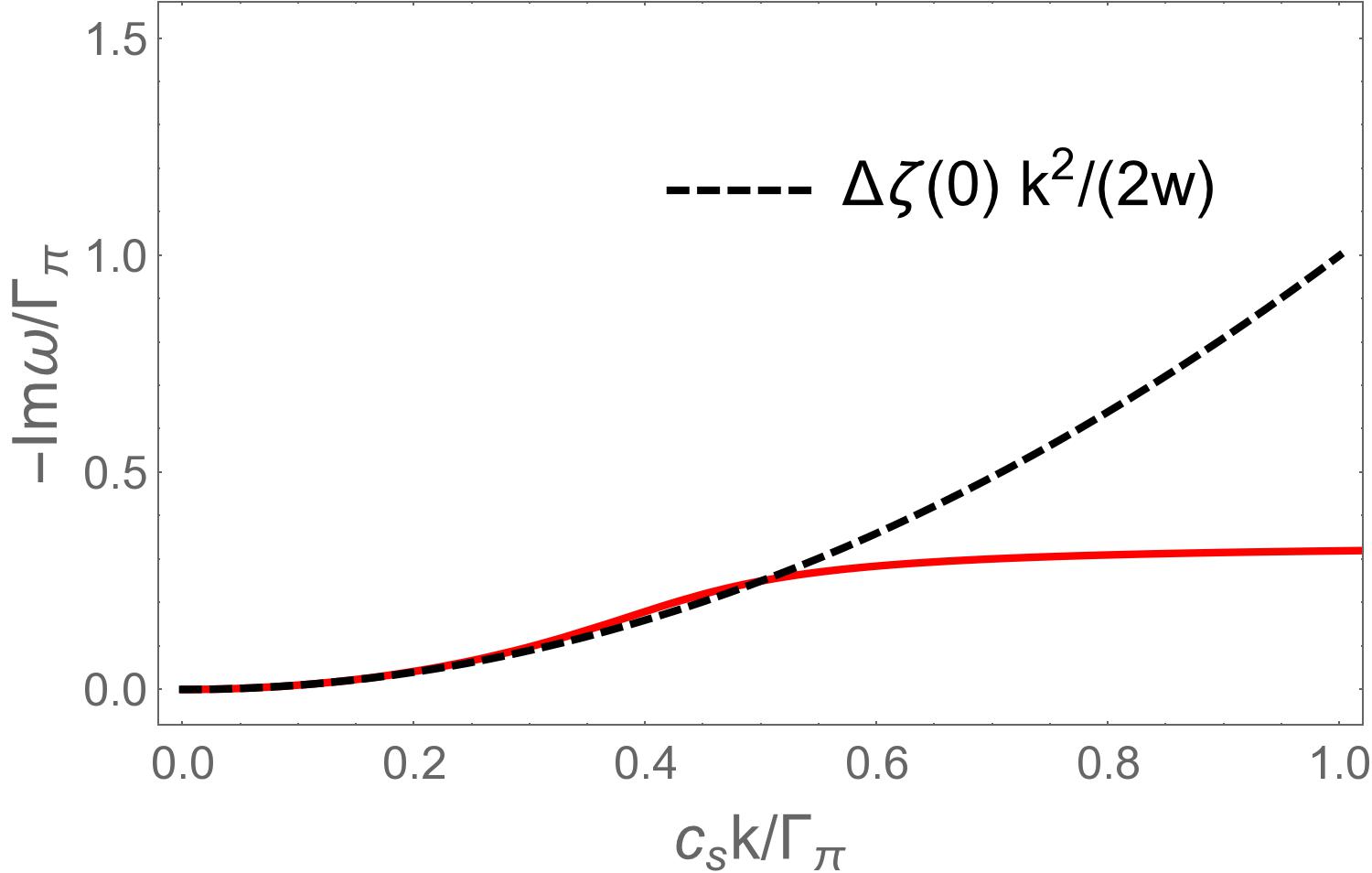}
\label{fig:disIm}
}
\subfigure[]
{
\includegraphics[width=.45\textwidth]{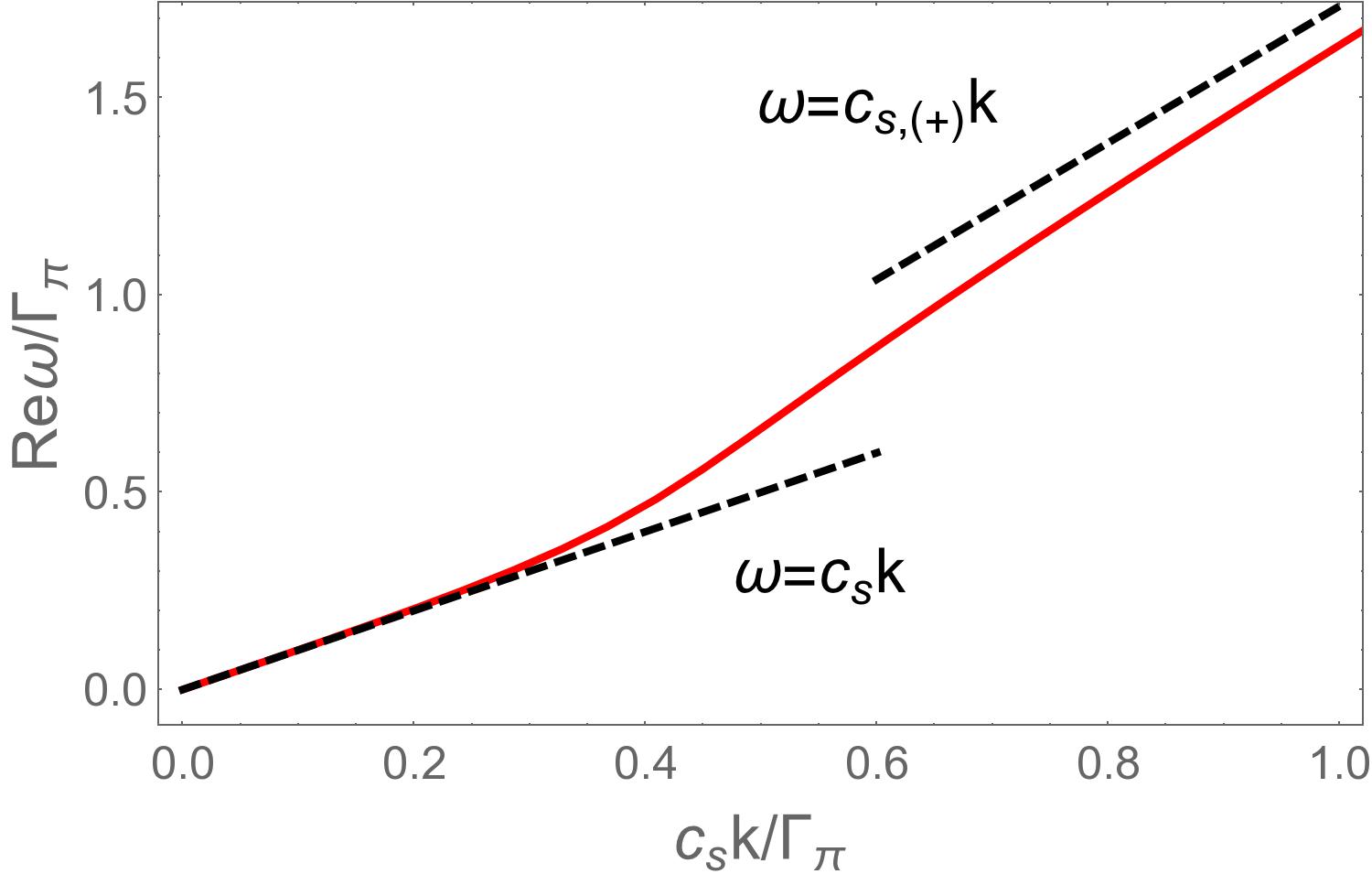}
\label{fig:disRe}
}
\caption{
 (color online) Sound dispersion relation, i.e., real (right) and
 imaginary (left) part of sound
 frequency as a function of $k$, in single-mode Hydro+
 determined by solving \eqref{eq:disp-rel}, compared to ordinary
 hydrodynamics (dashed lines for $\omega\ll\Gamma_\pi$). The quantities are normalized to make the plots
 scale-independent. The dimensionless ratio $\D
 c^{2}_{s}/c^{2}_{s}$ is set to $2$ for concreteness. 
  }
\label{fig:dis}
\end{figure}

Substituting the pressure oscillation given by
Eq.~(\ref{eq:delta-p-plus}) into the linearized hydrodynamic equations
we find the dispersion relation for the sound as well as the
non-hydrodynamic slow mode given by the three solutions of
\begin{equation}
  \label{eq:disp-rel}
  \omega^2 = k^2 \left(
c_s^2 + 
\frac{\omega}{\omega +i \Gamma_\pi}\,\Delta c_s^2(\infty)
\right)
\end{equation}
In Fig.~\ref{fig:dis} the real and imaginary parts of the sound dispersion
relation given by Eq.~(\ref{eq:disp-rel}) are plotted for
illustration. Note that the behavior of sound attenuation rate changes
from quadratic in the regime $\omega\ll\Gamma_\pi$ to a constant for
$\omega\gg\Gamma_\pi$:
 \begin{align}
%\begin{equation}
  \label{eq:attenuation}
& \omega\ll\Gamma_\pi:\qquad 
{\rm Im}\, \omega = -\frac{k^2}{2} \frac{\Delta c_s^2(\infty)}{\Gamma_\pi}
= -\frac{k^2}{2} \frac{\Delta\zeta(0)}{w}
\\
%\end{equation}
%\begin{equation}
  \label{eq:attenuation-large-omega}
&  \omega\gg\Gamma_\pi:\qquad
{\rm Im}\,\omega = -\frac{\Gamma_\pi}{2}\frac{\Delta c_s^2(\infty)}{{c_s^2}_\plus}
%\end{equation}
   \end{align}
For even larger $\omega$ one also has to take into account the usual $\mathcal O(k^2)$
contribution unrelated to the slow mode coming from $\zeta_\plus$ (as well as
as from $\eta_\plus$ and $\lambda_\plus$).

To summarize this section we have considered the response of a system
with a parametrically slow mode, described by Hydro+, to bulk
expansion/compression. Most notably, for frequencies
$\omega\gtrsim\Gamma_\pi$ the effective stiffness
$\delta p/\delta\ed$, or the sound speed, increases (see
Figs.~\ref{fig:ReG} and~\ref{fig:disRe}), while the
frequency-dependent bulk viscosity drops (see
Fig.~\ref{fig:ImG}). Note that without this drop the sound
attenuation rate in Eq.~(\ref{eq:attenuation}) would have overcome the
sound frequency (compare dashed lines on Figs.~\ref{fig:disIm}
and~\ref{fig:disRe}). Instead, the sound attenuation rate (rather than
growing as $k^2$) saturates at a constant (Fig.~\ref{fig:disIm}). Such
frequency and wave-vector dependence is beyond the reach of
conventional hydrodynamics.

We want to emphasize again that these results, for
$\omega\gtrsim\Gamma_\pi$, are reliable if the slow mode $\phi$ is
{\em parametrically} slower than all the other non-hydrodynamic
modes. I.e., if $\Gamma_\pi$ is much smaller than the microscopic
(non-hydrodynamic) relaxation rates. This is an essential condition
which distinguishes Hydro+ from other descriptions which add
degrees of freedom not parametrically separated from other
microscopic modes, e.g., conventional Israel-Stewart hydrodynamics
\cite{Israel:1979wp} (see, e.g., Ref.~\cite{Geroch:2001xs} for a
discussion of the (in)applicability of Israel-Stewart theory.)
If, however, one treats the relaxation time $\tau_\Pi$ of the trace of
the stress tensor as a parameter which can be made arbitrarily large,
then an Israel-Stewart theory becomes an example of a single-mode
Hydro+ (with $p_\pi=1$). As expected, bulk viscosity diverges as
$\tau_\Pi\to\infty$ in such a theory~\cite{Monnai:2016kud}.

 %=================================
 %
 %   Plots of Sound dispersion
 %
%=================================

\section{Entropy of fluctuations and 2PI action}
\label{sec:entr-fluct-2pi}

\subsection{Introduction}

The purpose of this section is to introduce a particular new set of
degrees of freedom into hydrodynamics. Unlike the usual hydrodynamic
degrees of freedom, which describe local averages of conserved
densities, these additional degrees of freedom describe fluctuations
(and their correlations). This section explains the conceptual
framework which makes this possible, i.e., explains how one needs to
think about these degrees of freedom and the type of states of the
system that they describe.  Our limited goal here is to make this
understanding quantitative enough that we can determine the entropy
associated with these additional degrees of freedom, i.e.,
$s_{(+)}$. However, the conceptual framework we lay out is general and
broad and could be extended to other interesting problems (e.g.,
non-gaussian fluctuations).

We shall attempt to answer the question: what is the physical meaning
of a state characterized by given magnitudes of fluctuations and
correlations which are not equal to equilibrium values? What we need
to know is how much entropy the state with non-equilibrium
fluctuations/correlations is missing (because we ``know'' more about
it) compared to complete-equilibrium state. The answer to that
question will be given by Eq.~(\ref{eq:S2-S1}) at the end of this
section. The result and the formalism leading to it is remarkably
similar to the formalism of the 2-particle irreducible (2PI) effective
action in quantum field
theory~\cite{Luttinger:1960ua,Baym:1962sx,deDominicis:1964zz,Cornwall:1974vz,Norton:1974bm}.
The similarity is mathematical, while the physical origin and meaning
is different.\footnote{The 2PI action formalism has been used before
  to describe non-equilibrium evolution of quantum
  systems~\cite{Berges:2003pc}. Although still different, this
  application of 2PI bears the closest similarity to our approach. We
  hope that our discussion will provide an intuitive insight into the
  meaning of this formalism as well.}

The purpose of this section is to derive Eq.~(\ref{eq:S2-S1}) in such
a way as to elucidate its conceptual meaning in the present context.
A reader who finds Eq.~(\ref{eq:S2-S1}) and its physical meaning
evident following the intuitive explanation in the paragraph below it
may skip directly to Eq.~(\ref{eq:S2-S1}) on first reading. Another
important result in this section is Eq.~(\ref{eq:4}) which describes
``renormalization'' of the equation of state by fluctuations.

The discussion will be pedagogical and self-contained. We shall begin
with a standard textbook introduction (e.g., Ref.~\cite{Lifshitz:v5})
to statistical mechanics and the concept of local thermodynamic
equilibrium in order to emphasize the points which will be essential
for our purpose.

The most important point to keep in mind is that a {\em state} in
statistical physics describes an {\em ensemble} of microscopic quantum
states of the system. The concept reflects the impracticality of
describing the evolution of a macroscopic system by specifying and
evolving its (highly excited) microscopic state. Not only is such a
description unworkable, it is also unrealistic when we have no
practical way to prepare such a pure microscopic state for a macroscopic
system.

Instead, the statistical description deals with the ensemble of
``similar'' microscopic states. The ensemble is characterized by the
set of {\em probabilities} of each state to be in the ensemble (the density
matrix), or alternatively, the set of correlation functions, or
expectation values of operators in the ensemble. The space of all
possible  statistical states is thus  greater
than  the Hilbert space of microscopic (pure) states.

\subsection{Equilibrium fluctuations}

{\em Equilibrium states} form a special class of the statistical
states in which the probability of a microscopic state to appear in
the ensemble is a function {\em only} of conserved quantities (quantum
numbers) of this state, such as energy, charge, momentum, etc. An
equilibrium state is a very good approximation to a macroscopic system
(i.e., a system with many degrees of freedom) averaged over time
intervals sufficiently long compared to microscopic time scales (e.g.,
mean free time between collisions in a gas). Below we shall denote the
set of the conserved quantities by a vector $\Psi$.

As a warm-up let us first consider the simplest equilibrium
statistical state -- the microcanonical ensemble -- where the only
microscopic quantum states in the ensemble are those with $\Psi$
within a small interval $\Delta\Psi$ around a given value
$\bar\Psi$. The size of the interval $\Delta\Psi$ is assumed to be
small compared to characteristic scale of the variation of the density
of states, but much larger than the level spacing. For a macroscopic
system, with very large density of states, this is easily
satisfied. The probability of each state in this interval is the
same. Then entropy, which we shall denote $S_0(\bar\Psi)$, is given
simply by the logarithm of the number of the states in the ensemble.

Next let us consider canonical ensemble -- an open system in contact
with a much larger (infinite) reservoir of conserved quantities
$\Psi$. All states of the system can now appear, but the probability
of each state is now weighed by the exponential $\exp(J\Psi)$, where
$J$ is a set (a vector) of thermodynamic variables conjugate to $\Psi$
(e.g., $\mu/T$ for charge, or $-1/T$ for energy):
$  J = -{\partial S_\textrm{reservoir}}/{\partial\Psi} $, and we use symbolic notation
$J\Psi=\sum_n J_n \Psi_n$ for the sum of the products of each variable
and its conjugate. % {\sf (Change sign of J and K?)}
 This extra weight
reflects the dependence of the number of states of the reservoir when
it exchanges energy with our system. 

The entropy of the canonical ensemble can be calculated by using the standard
formula:
\begin{equation}
  \label{eq:S}
  S = -\sum_i p_i \log p_i
\end{equation}
where the sum runs over all states and $p_i$ is the probability of a
state labeled $i$. Since this is an equilibrium state (by definition)
the value of $p_i$ depends only on the value $\Psi$ for this
state. The probability of each state is the same up to a factor
$e^{J\Psi}$. Since the
density of states is large we can replace the sum over $i$ with the
integral over $\Psi$, taking into account the density of states:
\begin{equation}
\sum_i \to \int (d\Psi/\Delta\Psi) e^{S_0}. \label{eq:sum-int}
\end{equation}
The
{\em normalized\/} probability for each microscopic state is given by
\begin{equation}
  \label{eq:p_i}
  p_i = \exp(J\Psi-W(J))
\end{equation}
where
\begin{equation}
  \label{eq:W}
  e^{W(J)} = \int (d\Psi/\Delta\Psi) e^{S_0(\Psi)+J\Psi}
\end{equation}
The entropy of the canonical ensemble in Eq.~(\ref{eq:S}) is then
given by
\begin{equation}
  \label{eq:SWJ}
  S = - \int d\Psi e^{S_0(\Psi)+J\Psi-W(J)}(J\Psi - W(J)) = W(J) - J\langle\Psi\rangle
\end{equation}
Since $\langle\Psi\rangle = dW/dJ$, $S(\langle\Psi\rangle)$ is a Legendre transform of $W(J)$.

\subsection{Partial-equilibrium fluctuations}

As the next step let us consider so-called {\em partial} (or,
incomplete) equilibrium states~\cite{Lifshitz:v5}, i.e., the states
where the equilibrium has been achieved locally, for regions of size
of order, say, $\ell$, much larger than the microscopic scale (such as
mean free path), but the global (complete) equilibrium has not yet
been achieved. This is a common situation when the system is observed
over {\em finite} periods of time because the relaxation time grows with the
size of the system. The evolution of these states is described by
hydrodynamics. We can generalize the calculation of entropy for these
states by using additivity of entropy. Each of the subsystems of size
$\ell$ can be treated as an open system in equilibrium with its local
environment and its entropy calculated using the formulas for
a canonical ensemble above. The total entropy is then the sum of the
entropies of the parts. More formally, this means considering
microscopic state probabilities $p_i$ as functionals of the (slowly
varying on the scale $\ell$) fields $\Psi$, and replacing the
integrals in Eq.~(\ref{eq:SWJ}) with the path integrals. We thus find
the entropy of a partial equilibrium state as a functional of slowly
varying $\bar\Psi\equiv\langle\Psi\rangle$:
\begin{equation}
  \label{eq:S1}
    S_1[\bar\Psi] = - \int \mathcal D \Psi e^{S_0[\Psi]+J\Psi-W[J]}(J\Psi - W[J]) = W[J] - J\langle\Psi\rangle
\end{equation}
where $J\Psi$ now denotes the sum over the variables as before as well as the
integral over space (i.e., sum over the locally equilibrated
subsystems): $J\Psi = \sum_n\int_{\bm x} J_n(\bm x) \Psi_n(\bm x)$.
The entropy of a partial equilibrium state is thus given by the Legendre
transform of the ``one-particle irreducible'' (1PI) generating functional $W[J]$ of the correlation
functions of $\Psi$.

Finally, we must realize that the partial-equilibrium state we described
above is still a special case and in many important situations is
inadequate for describing a real system out of equilibrium even on the
time scales sufficient for local equilibration.  Since only local
equilibration had enough time to occur, the profile of $\Psi$ in the
system on scales longer than $\ell$ is still different from
equilibrium (constant in space) value. Also, each local value of
$\bar\Psi$ is different in different members of the ensemble. In the
states we just described these fluctuations within the ensemble are,
however, completely determined by $\bar\Psi$ (or, alternatively, $J$)
and the density of states. This property of the ensemble is, however,
unrealistic in many important cases when not only $\bar\Psi$, but also
its variations in space on scales longer than $\ell$ between different
members of the ensemble are not what they should be in
equilibrium. The relaxation time of such long-wavelength fluctuations
is of the same order as the relaxation time for long-wavelength inhomogeneities
of $\bar\Psi$ itself and thus these fluctuations would typically be out
of equilibrium at these time scales.

In other words, we must consider states where not only one-point
function $\bar\Psi=\langle\Psi\rangle$, but also the two-point
functions $\langle\Psi_n(\bm x)\Psi_m(\bm y)\rangle$), which we shall
denote symbolically as $\langle\Psi\Psi\rangle$, and, in general,
higher order correlation functions are still away from equilibrium
values. In other words, the state should be characterized by the
probability {\em functional} $p_i[\Psi]$, which is not completely determined
by the average value $\bar\Psi$, but can be an arbitrary functional,
which will evolve in time as it approaches the equilibrium form.

To formalize this observation, we recall that the equilibrium probability of
the microscopic state we considered is given by
equation~(\ref{eq:p_i}). To allow for a state with arbitrary value of
a two-point function we can consider the probability in the form:
\begin{equation}
  \label{eq:p_iK}
  p_i = \exp(J\Psi + \frac12\Psi K \Psi - W_2[J,K])
\end{equation}
where we introduced an arbitrary quadratic form matrix/operator $K$ to
parameterize the deviation of the probability distribution from
equilibrium. The normalization of the probability is given by
\begin{equation}
  \label{eq:W2}
  e^{W_2[J,K]} =   \int \mathcal D \Psi
  e^{S_0[\Psi]+J\Psi+\frac12\Psi K\Psi}
\end{equation}
where $\Psi K\Psi$ denotes $\int_{\bm x\bm y}\sum_{nm}\Psi_n(\bm x)K_{nm}(\bm
x,\bm y)\Psi_m(\bm y)$.
The entropy of such a partial equilibrium state is given by the standard formula
Eq.~(\ref{eq:S}):
\begin{multline}
  \label{eq:S2}
  S_2 = -\sum_i p_i\log p_i = 
\int \mathcal D\Psi
e^{S_0[\Psi]+J\Psi+\frac12\Psi K\Psi-W_2[J,K]}(J\Psi +\frac12\Psi
K\Psi- W_2[J,K]) 
\\= W_2[J,K] - J\langle\Psi\rangle - \frac12\langle\Psi
K\Psi\rangle
\end{multline}
We find that the entropy is the Legendre transform of the 2PI
generating functional $W_2[J,K]$. It should be possible to generalize
this to higher point correlators, and this should be important to do
in order to study higher
moments of fluctuations near the critical point. We defer this to
further work.

Here we focus on states described by one- and two-point functions. It
is convenient to introduce a correlator 
\begin{equation}
  \label{eq:G}
  G = \langle\Psi\Psi\rangle - \bar\Psi\bar\Psi
\end{equation}
in terms of which
\begin{equation}
  \label{eq:KG}
  \langle \Psi K\Psi\rangle = \tr KG + \bar\Psi K \bar\Psi
\end{equation}
 It is
convenient to substitute Eq.~(\ref{eq:KG}) into Eq.~(\ref{eq:S2}):
\begin{equation}
  \label{eq:S2-G}
  S_2 = W_2[J,K] - J\langle\Psi\rangle - \frac12\bar\Psi K \bar\Psi 
- \frac12 \tr KG
\end{equation}
and express $S_2$ as a functional of $\bar\Psi$ and $G$ using implicit
equations
\begin{equation}
  \label{eq:PsiG}
  \bar\Psi = \frac{ \delta W_2}{\delta J}
\qquad\mbox{and}\qquad
G + \bar\Psi\bar\Psi = 2\frac{\delta W_2}{\delta K}
\end{equation}

It is useful to note that the derivatives of $S_2[\bar\Psi, G]$ are
given by
\begin{equation}
  \label{eq:-JK}
   \frac{ \delta S_2}{\delta \bar\Psi} = -J
\qquad\mbox{and}\qquad
2\frac{\delta S_2}{\delta G} = -K
\end{equation}
Therefore, one can think of $K$ as a thermodynamic restoring
force bringing the system back to equilibrium.

To be more explicit, we shall evaluate $W_2$ in Eq.~(\ref{eq:W2}) in
the saddle point approximation (which should be appropriate in the
regime $\ell\gg\xi$ due to central limit theorem):
\begin{equation}
  \label{eq:W2sp}
  W_2[J,K] \approx S_0[\bar\Psi] + J\bar\Psi + \frac12\bar\Psi K \bar\Psi
- \frac12\log\det(C-K)\,,
\end{equation}
where we introduced quadratic form matrix/operator 
\begin{equation}
  \label{eq:C}
  C = -\frac{\delta^2 S_0}{\delta\bar\Psi\delta\bar\Psi}\,.
\end{equation}
The value of $\bar\Psi$ is determined by $J$ and $K$ via
the saddle-point equation
\begin{equation}
  \label{eq:saddle}
  \frac{\delta S_0}{\delta\bar\Psi} + J + K\bar\Psi = 0.
\end{equation}

Substituting Eq.~(\ref{eq:W2sp}) into Eq.~(\ref{eq:S2}) and using
(\ref{eq:KG}) we find
\begin{equation}
  \label{eq:S_2sp}
  S_2[\bar\Psi,G] \approx S_0[\bar\Psi] - \frac12\tr KG - \frac12\log\det(C-K)
\end{equation}
To eliminate $K$ in favor of $G$ we apply Eq.~(\ref{eq:PsiG}) to
Eq.~(\ref{eq:S2-G}) to find
\begin{equation}
  \label{eq:GK}
  G = (C-K)^{-1} 
\end{equation}
which we substitute into Eq.~(\ref{eq:GK})
\begin{equation}
  \label{eq:S2-S0}
  S_2[\bar\Psi,G] \approx S_0[\bar\Psi] - \frac12 \tr(CG-1) +
  \frac12\log\det G.
\end{equation}
It is more convenient to express microscopic action $S_0$ in terms of
the 1PI effective action $S_1$ defined in Eq.~(\ref{eq:S1}). It is
easy to see that $S_1=S_2|_{K=0}$, i.e.
\begin{equation}
  \label{eq:4}
  S_1\approx S_0 + \frac12\log\det C
\end{equation}
and thus
\begin{equation}
  \label{eq:S2-S1}
  S_2[\bar\Psi,G] \approx S_1[\bar\Psi] - \frac12 \tr(CG-1) +
  \frac12\log\det CG.
\end{equation}

One can see that $S_2\le S_1$. The entropy is maximized when
$G=C^{-1}$, i.e., when the fluctuations (characterized by the 2-point
function $G$) are in equilibrium ($K=0$).

To understand this result physically it is useful to keep in mind that
entropy is a measure of the uncertainty of the system. Since larger
fluctuations mean larger uncertainty, the entropy should increase with
$G$. The last, logarithmic, term in Eq.~(\ref{eq:S2-S1}) describes
this. One can think of $\sqrt{\det G}$ as a measure of the ``spread''
of the thermodynamic state probability distribution over the
microscopic states, and the logarithm of it is the entropy. However,
the increase in the magnitude of fluctuations around equilibrium comes
at an expense: due to the convexity of the entropy the average entropy
of a state decreases when the fluctuations get larger, i.e.,
$\langle S\rangle=S_0 +
\delta^2S/(\delta\Psi\delta\Psi)\langle\delta\Psi\delta\Psi\rangle<S_0$.
This effect is described by the second term in
Eq.~(\ref{eq:S2-S1}). The balance of these two effects leads to the
maximum at the equilibrium value of the fluctuations given by
$G=C^{-1}$.

\subsection{Scale separation and mode distribution function}

Let us consider a system in a partial-equilibrium state where the
equilibrium is ``complete'' only for subsystems of a macroscopic size
$\ell$ (or smaller). If $\ell\gg\xi$, in such a state the contribution
of fluctuations is suppressed according to the central limit
theorem,
% . The suppression is controlled by a small parameter
% $(\xi/\ell)^{3/2}$
where $\xi$ is the correlation length of
fluctuations. This separation of scales $\ell\gg\xi$ is also reflected
in the dependence of the fluctuation correlator $\bar G(\bm x_1,\bm
x_2) = C^{-1}(\bm x_1,\bm x_2)$ on $\vx_1$ and $\vx_2$. Indeed, in the
fully equilibrated state ($\ell=\infty$), which is homogeneous, the
local value of $\Psi$ is position-independent and the  two-point
correlator, $\bar G(\bm x_1,\bm x_2) = C^{-1}(\bm x_1,\bm x_2)$,
depends only on the difference $\bm x_1 - \bm
x_2$. %, and not on $(\bm x_1 +\bm x_2)/2$.
The typical scale for this dependence is $|\vx_1-\vx_2|\sim\xi$.
In the partial equilibrium
states $\bar\Psi(\bm x)$ depends on $\bm x$ very slowly and similarly
the dependence of $\bar G(\bm x_1,\bm x_2)$ on $(\bm x_1+\bm x_2)/2$ is slow,
and is associated with the scales longer than $\ell$ which is much
longer than the scale $\xi$ of $|\vx_1-\vx_2|$ dependence.

 This separation of scales is most conveniently exploited by
performing Wigner transform of $G$, i.e., Fourier transform
w.r.t. $\bm x_1 - \bm x_2$.
\begin{equation}
  \label{eq:GW}
  G_{\bm Q}(\bm x) 
= \int_{\Delta \bm x}
G\left(\bm x+\frac{\Delta\bm x}{2},\bm x- \frac{\Delta\bm x}{2}\right) 
e^{i\bm Q \Delta\bm x}.
\end{equation}
The states we consider are characterized by $G_{\bm Q}$ which vary slowly
with $\bm x$, compared to the scale set by characteristic value of $Q$.

For such states the 2PI action simplifies. The functional trace in
Eq.~(\ref{eq:S2-S1}) becomes an integral over $\bm x$ and over $\bm Q$
of a matrix function of $G_{\bm Q}(\bm x)$, i.e,
\begin{equation}
  \label{eq:S2-S1-W}
  S_2[\bar\Psi,G] \approx S_1[\bar\Psi] 
+ \frac12\int_{\bm x}\int_{\bm Q}{\rm Tr}\,\left(1-C_{\bm Q}G_{\bm Q} +
  \log C_{\bm Q}G_{\bm Q}\right).
\end{equation}
where $C_{\bm Q}=\bar G_{\bm Q}^{-1}$ is the Wigner transform of $C$
from Eq.~(\ref{eq:C}).  We use intuitive short-hand notations for spatial and
wave-vector integrations respectively:
\begin{equation}
  \label{eq:intQ}
 \int_{\vx} \ldots = \int d^3\vx \ldots
 \quad\mbox{and}\quad 
\int_\vQ \ldots \equiv \int d^3\left(\vQ/2\pi\right)\ldots\,.
\end{equation}

It is notable that a similar scale separation occurs in kinetic theory
where the particle distribution function is also a Wigner transform of
a two-point correlator. 
Mathematical similarity notwithstanding, the
physical origin of the separation is different in that case -- the
slowness of the collision rate compared to typical particle momenta.
In our case such a Wigner transform would be more appropriately called
a {\em mode} distribution function, since the variable $\bm Q$ is a
wave-vector of the mode and not a particle momentum.
The integral over variables ${\bm x}$ and ${\bm Q}$ is
the phase-space integral. The evolution equation for the mode
distribution function is similar to a kinetic equation, and one could
use this term to describe it, with the understanding that it does not
describe particle kinetics, but rather the mode kinetics.\footnote{For
the sound channel the similarity is not just mathematical, but also
physical: The corresponding matrix element of $G_{\bm Q}$ can
be identified with the distribution function of phonon quasiparticles,
and the corresponding equation with the kinetic (Boltzmann) equation
for the phonon quasiparticles as in Ref.~\cite{andreev1970twoliquid}.}

\section{Kinetics of fluctuations}
\label{sec:fluctuation-modes}

\subsection{Relaxation equations}

The evolution of the system we are describing is governed by the
second law of thermodynamics, i.e., the evolution proceeds towards
the maximum of the entropy under (energy-momentum, charge, etc)
conservation constraints. Hydrodynamic equations are essentially these
constraints (supplemented with constitutive equations). We can use the
same approach to describe the evolution of the 2-point (as well as
1-point) functions towards the maximum of the 2PI entropy $S_2$. In
order to do that, we need to supplement the usual set of hydrodynamic
equations for the conserved densities $\bar\Psi$, with the equations
which describe the time evolution of the 2-point functions
$G$. Following the same logic, we can write these equations as
relaxation equations. Since we focus on the quadratic fluctuations, we
would need linearized form of the hydrodynamic equations. Writing the
hydrodynamic equations (in matrix notations) in the Onsager form and
linearizing them we find
\begin{equation}
  \label{eq:dtPsi}
  D \Psi = -\gamma J = - L \delta \Psi + \mathcal O(\delta\Psi^2)
\end{equation}
where $\gamma$ is the Onsager matrix and $L$ is the linear evolution
matrix which are related via
\begin{equation}
  \label{eq:gamma=LC-1}
  \gamma = L C^{-1}\,.
\end{equation}
The explicit form of this matrices is presented in Section~\ref{sec:slowest-mode}.

 We can now write the
linearized evolution equation for the 2-point functions in matrix notations as
\begin{equation}
  \label{eq:dtG}
  \partial_t G =  - L(G-\bar G) - (G-\bar G)L^\dag + \mathcal O(G-\bar G)^2 
\end{equation}
where $\bar G=C^{-1}$ is the equilibrium value of $G$. This equation
is easy to derive from equation~(\ref{eq:dtPsi}) with a noise
term. 

In order to determine the full non-linear form of Eq.~(\ref{eq:dtG})
we need to use the expression for the entropy
Eq.~(\ref{eq:S2-S1}) and the second law of thermodynamics.  In order
for the law to hold we need the relaxation equation to have the
Onsager form, i.e., the relaxation rate should be proportional to the
thermodynamic force
\begin{equation}
  \label{eq:piG}
  K = - 2\frac{\delta S_2}{\delta G} 
= -\left(G^{-1}-\bar G^{-1}\right)
= C\left(G-\bar G\right)C + \mathcal O(G-\bar G)^2 
\end{equation}
where we used $C=\bar G^{-1}$ (factor of 2 is due to symmetry of
$G$). It is easy to see that the equation
which obeys this condition and agrees with Eq.~(\ref{eq:dtG}) has the
form:
\begin{equation}
  \label{eq:dtG-nl}
  \partial_t G = - \gamma K C^{-1} -  C^{-1} K \gamma 
=  \gamma\left(G^{-1}\bar G-1\right) + \left(\bar G G^{-1}-1\right)\gamma\,.
\end{equation}

So far the discussion of the kinetics was general and did not assume
separation of scales, $Q\gg 1/\ell$, discussed in the previous
subsection. Such scale separation simplifies equations in terms of the
Wigner transformed functions. Symbolically the form of equations in
terms of $G_{\bm Q}$ remains the same, but we can replace all
matrix/operators ($G$, $C$, $L$, $\gamma$) by their Wigner
transforms evaluated at the common wave-vector $\bm Q$.

\subsection{Mode decomposition}

One can solve the linear equation (\ref{eq:dtPsi}) by decomposing the
set of variables $\Psi$ into the right eigenmodes of the operator
$L$. Then the solution is given by the sum of projectors
\begin{equation}
  \label{eq:LP}
  L = \sum_n \lambda_n P_n, \qquad \mbox{where } P_n\equiv\rv_n \lv_n^\dag
\end{equation}
and $\rv$ and $\lv$ are the right an left eigenmodes of $L$ respectively:
\begin{equation}
  \label{eq:Lrvlv}
  L\rv_m = \lambda_m\rv_m,\qquad \lv_n^\dag L = \lambda_n\lv_n^\dag
\end{equation}
Although the right and left eigen{\em{values}} are the same (roots of the
characteristic polynomial), the right and left eigen{\em{vectors}} are, in
general, different (i.e., $P_n^\dag\neq P_n$) and not orthogonal among
themselves. Instead they form a set of dual bases which are {\em mutually\/}
orthogonal, i.e.,
\begin{equation}
  \label{eq:rv-orth}
  \lv_n^\dag \rv_m = \delta_{mn},\qquad\mbox{or } P_n P_m = \delta_{mn}P_n\,.
\end{equation}
One can show this by multiplying the first and the second of
equations~(\ref{eq:Lrvlv}) by $\lv_n^\dag$ on the left and by $\rv_n$
on the right respectively. 

Although $L$ is not hermitian, the Onsager matrix/operator in
Eq.~(\ref{eq:gamma=LC-1}) is, which means
\begin{equation}
  \label{eq:LC}
  LC^{-1} = C^{-1} L^\dag\,.
\end{equation}
Therefore matrices $C$ and $C^{-1}$ (which are hermitian) can be written as
\begin{equation}
  \label{eq:CC-1}
  C = \sum_n c_n \lv_n\lv_n^\dag,
\qquad
C^{-1} = \sum_n c_n^{-1} \rv_n\rv_n^\dag\,.
\end{equation}
The result for $C^{-1}$ can be derived by ``sandwiching''
Eq.~(\ref{eq:LC}) between $\lv_m^\dag$ and $\lv_n$ and using
Eq.~(\ref{eq:Lrvlv}) and Eq.~(\ref{eq:rv-orth}), while the result for
$C$ -- by doing the same to equation $CL=L^\dag C$ using $\rv_m^\dag$ and
$\rv_n$.
\footnote{The coefficients $c_n$ can be chosen arbitrarily by
  adjusting the normalization of $\rv_n$ and $\lv_n$, while still
  preserving Eqs.~(\ref{eq:rv-orth}). One could, for example, choose
  $c_n\equiv 1$. However, there are other considerations which make
 certain other choices preferable. }

If the matrix $L$ were hermitian, this would reduce to a familiar
result that $C$ can be diagonalized in the same basis as $L$ since
Eq.~(\ref{eq:LC}) would become commutativity condition.

\subsection{Projection onto the slowest mode}

Our discussion of fluctuations in this section is general. In 
Section~\ref{sec:hydro+-near-qcd}
 we shall consider a special case where a {\em parametric}
separation of scales appears between the relaxation rates of different
modes of fluctuations, such as the case near a critical point. The
slowest relaxing mode in this case is the heat diffusion at constant
pressure (see Section~\ref{sec:slowest-mode}), whose relaxation time $\ell^2/D_p$ is
longest because the diffusion coefficient $D_p$ vanishes at the
critical point. This is due to the divergence of heat capacity $c_p\sim\xi^2$
and the relation $D_p=\kappa/c_p$, where
$\kappa = \lambda\,(\beta w/n)^2$ is the heat conductivity. As a
result $D_p\sim \xi^{-1}$, even despite the divergence of $\lambda\sim\xi$ (we round all powers of $\xi$ to integer
values for simplicity).

In such a situation one may consider a partial-equilibrium state where
the ``complete'' (local) equilibrium of all modes is achieved on
length scales $\ell$, except for the slowest mode, whose equilibrium
still needs more time to be reached. In this case we can neglect the
fluctuations of all equilibrated modes since their contribution is
typically suppressed by
central limit theorem by a factor $(\xi/\ell)^3\ll1$.\footnote{There are, of course, special measurements where
  such fluctuations give leading contributions, e.g., in long-time
  tails of correlators \cite{andreev1970twoliquid,Kovtun:2003vj}. In this case,
  since $\ell^2\sim t$, the suppression factor
  $(\xi/\ell)^3\sim t^{-3/2}$ leads to the characteristic
  half-integer power tail.} The
unequlibrated fluctuation mode can then be treated using the formalism
we introduce in the next section.

The slowest mode of equation (\ref{eq:dtPsi}), $\psi_1$, corresponds to the
smallest (in terms of its real part)
eigenvalue of $L$, $\lambda_1$.\footnote{It is also useful to note
  that due to $L=\gamma C$ %and Eq.~(\ref{eq:CC-1}) 
that mode is also
  the flattest direction of the quadratic form~$C$. A simple explicit
  example of this could be found in Ref.\cite{Son:2004iv}.}
 I.e., the slowest
mode is the projection $P_1\Psi=\rv_1(\lv_1^\dag\Psi)$. It is easy to see
that the relaxation rates in Eq.~(\ref{eq:dtG}) are given by
$\lambda_n+\lambda_m$, i.e., the slowest relaxation rate corresponds
to $n=m=1$ and the slowest mode is given by the projection
$P_1(G-C^{-1})P_1^\dag$. If we neglect all other (faster) modes, the matrix $G$ will take the
form:
\begin{equation}
  \label{eq:GPhi}
  G = C^{-1} + P_1 (G -C^{-1}) P_1^\dag = C^{-1} + (\phi-c_1^{-1})\rv_1\rv_1^\dag
  = \left(1 + (\phi c_1-1)P_1\right)C^{-1}
\end{equation}
where we introduced
\begin{equation}
  \label{eq:Phi0G1}
  \phi \equiv \lv_1^\dag G\lv_1
\end{equation}
and used Eq.~(\ref{eq:CC-1}).

Since $G$ is not just a discrete matrix, but an operator whose kernel
$G(\bm x, \bm y)$ is a $5\times5$ matrix, the spectrum of modes is not discrete,
but a continuous spectrum of hydrodynamic modes. For the
partial equilibrium states we consider, where the rate of variation
w.r.t. $\bm x + \bm y$ is much slower than w.r.t. $\bm x - \bm y$,
i.e., $1/\ell\ll Q$, the problem
simplifies, as we have seen in Eq.~(\ref{eq:S2-S1-W}), where the
action can be written as a local functional of the Wigner transform
$G_{\bm Q}(\bm x)$. We can then consider the lowest eigenmode of  $G_{\bm
  Q}$ locally and define the corresponding projection
\begin{equation}
  \label{eq:PhiG1}
  \phi_{\bm Q}(\bm x) \equiv \lv_1^\dag G_{\bm Q}(\bm x)\lv_1
\end{equation}
which is related to $G_{\bm Q}$ as $\phi$ is related to $G$ in
Eq.~(\ref{eq:GPhi}). 
Substituting this expression for $G_{\bm Q}$  into
the 2PI entropy in Eq.~(\ref{eq:S2-S1-W}) we find
\begin{equation}
  \label{eq:S2-S1-Phi}
  S_2[\bar\Psi,G] \approx S_1[\bar\Psi] 
+ \frac12 \int_{\bm x}\int_{\bm Q}\left(1-\phi_{\bm Q}/\bar\phi_{\bm Q}  +
  \log (\phi_{\bm Q}/\bar\phi_{\bm Q})\right)
%- \frac12 (\Phi/\bar\Phi - 1) +  \frac12\log (\Phi/\bar\Phi)
\end{equation}
where 
\begin{equation}
  \label{eq:Phibar}
  \bar\phi_{\bm Q}(\bm x) \equiv c_1^{-1} = \lv_1^\dag \bar G_{\bm Q} \lv_1\,.
\end{equation}
Here $\phi_{\bm Q}(\bm x)$ is an additional degree of freedom whose local equilibrium $\bar\phi_{\bm Q}$
depends on $\bm x$ via the dependence on local equilibrium value of
hydrodynamic variables (1-point functions)~$\bar\Psi$.

In order to write the kinetic equation of the slowest mode $\phi_{\bm
  Q}$ we would need to eliminate faster modes using the equations of
motion such as Eq.~(\ref{eq:dtG-nl}). For linear equations this would
simply amount to the projection we described above. 
However due to
non-linearities this procedure is more complicated and should essentially
capture the known physics of the ``mode-coupling'' \cite{PhysRev.166.89,PhysRevA.1.1750,PhysRevE.55.403}. 
Here we shall use the result of the
mode-coupling calculations in \cite{PhysRevA.1.1750,PhysRevE.55.403} to write the
resulting equation as
\begin{equation}
  \label{eq:dtPhi}
  D\phi_{\bm Q} = -\gamma_\pi(\bm Q)\pi_{\bm Q}\,,
\end{equation}
where
\begin{equation}
  \label{eq:piQ}
  \pi_{\bm Q} \equiv - \frac{\delta S_2}{\delta\phi_{\bm Q}} = 
\frac12 \( \bar\phi_{\bm Q}^{-1} - \phi_{\bm Q}^{-1}\)\,.
\end{equation}
The coefficient $\gamma_\pi(\vQ)$ can be related to the relaxation rate
$\G(\vQ)$ which is known from the mode-coupling calculations and
we shall discuss it below (see Eq.~(\ref{kawasaki})). We have also
omitted the $A_\phi(\vQ)$ term because we shall choose the slowest
mode to be $s/n$ (see Section~\ref{sec:slowest-mode}), which has zero compressibility, i.e.,
$D(s/n)=0\cdot\theta +\ldots\ $. We remind the reader that the choice is
arbitrary (see Appendix~\ref{sec:repar-covar-hydr}), and a more detailed calculation
would be needed to determine what the optimal choice of the slow mode
should be, and if that choice has nonzero $A_\phi$. In addition, one
should also consider Hydro+ terms with $\lambda_{\pi\pi}$ and
$\lambda_{\alpha\pi}$, but we shall defer this as well as a more
nuanced choice of $\phi_\vQ$ and the derivation of
Eq.~(\ref{eq:dtPhi}) to future work.

\subsection{The slowest mode}
\label{sec:slowest-mode}

For completeness, we present here an explicit form of the matrix operators
$L$, $C$ and $\gamma$ and identify the slowest mode (or more
precisely, the branch of modes).

Writing linearized hydrodynamics in coordinates $\delta\Psi=(w\delta \bm v(\bm
k),\delta\varepsilon(\bm k),\delta
n(\bm k))$ we find:
\begin{equation}
  \label{eq:L-hydro}
  L = 
  \begin{pmatrix}
    \mathbb V & i\bm k p_\varepsilon & i\bm k p_n \\
    1 & 0 & 0 \\
    n/w & \bm k^2\lambda\alpha_\varepsilon & \bm k^2 \lambda \alpha_n \\ 
  \end{pmatrix}
\end{equation}
where subscripted thermodynamic variables $p_\varepsilon$, $p_n$,
$\alpha_\varepsilon$, $\alpha_n$ denote
derivatives of the variable with respect to the subscript variable
($\varepsilon$ or $n$) while the other variable  is held fixed (e.g.,
$p_\ed\equiv(\pd p/\pd\ed)_n$) and
$\mathbb V=\eta \bm k^2 + \left(\zeta+ \frac13\eta\right)\bm
k\otimes\bm k$ is a matrix of viscous relaxation.
\begin{equation}
  \label{eq:C-hydro}
  C =
  \begin{pmatrix}
    \beta/w & 0 & 0 \\
    0 & \beta_\varepsilon & \beta_n \\
    0 & -\alpha_\varepsilon & -\alpha_n \\
  \end{pmatrix}
%+ \mathcal O (\bm k^2).
\end{equation}
which is symmetric by virtue of the Maxwell relation 
$\beta_n=-\alpha_\varepsilon$.
Therefore the Onsager matrix is given by
\begin{equation}
  \label{eq:Onsager-hydro}
  \gamma = LC^{-1} = \frac1\beta
  \begin{pmatrix}
    w \mathbb V & i\bm k w & i\bm k n \\
    -i\bm k w   &  0 &  0 \\
    -i\bm k n   &  0 &  \bm k^2 \beta\lambda \\
  \end{pmatrix}
\end{equation}
where we
performed some standard thermodynamic Jacobian calculus to simplify
the result.

To lowest order in $k$ (ideal hydrodynamics), the smallest eigenvalue
of  matrix $L$ is 0 and the corresponding (unnormalized) right and left
eigenvectors are:
\begin{equation}
  \label{eq:rvlv1}
  \psi_1 \sim \left( \bm 0, 1, -p_\ed/p_n\right)\,,
\quad
\theta_1 \sim \left(\bm 0, 1, -w/n\right)\,.
\end{equation}
Since $-p_\ed/p_n = (\partial n/\partial\ed)_p$, this simply means
that in the mode $\delta \Psi \sim \psi_1$ the $\delta\ed$ and
$\delta n$ fluctuations are such that $\delta p=0$, i.e., pressure does
not fluctuate. The projection on that mode from an arbitrary
fluctuation, i.e, $\theta_1\delta\Psi\sim \delta\ed - (w/n) \delta n$
is proportional to the fluctuation of $s/n$ since:
\begin{equation}
  \label{eq:n/s-en}
  \delta\(\frac sn\)= \frac{\beta}{n}\left(\delta\ed -
    \frac{w}{n}\delta n \right)\,.
\end{equation}
Thus we can describe the slowest mode as the diffusion of entropy per
baryon at fixed pressure.

Therefore, the variable $\phi_\vQ$ defined by projection on the
slowest mode in Eq.~(\ref{eq:PhiG1}) can be identified with the Wigner
transform of the correlator of $s/n$:
\begin{equation}
  \label{eq:phiQmm}
  \phi_{\bm Q}(\bm x) 
\sim \int_{\Delta \bm x} \left\langle
\delta m\left(\bm x+\frac{\Delta\bm x}{2}\right)
\delta m\left(\bm x- \frac{\Delta\bm x}{2}\right)
\right\rangle 
e^{i\bm Q \Delta\bm x}\,.
\end{equation}
where we defined
\begin{equation}
  \label{eq:msn}
  m\equiv \frac sn.
\end{equation}
The normalization of $\phi_\vQ$ is arbitrary, as can be seen from
the expression for the 2PI action in Eq.~(\ref{eq:S2-S1-Phi}), where
it cancels. Moreover, it is worth noting that any function of
$\phi_\vQ$ can be chosen to represent slow relaxation, due to
the reparameterization invariance of Hydro+. However,
one needs to be aware that the value of compressibility $A_\phi$ may
depend on that choice (see Appendix~\ref{sec:repar-covar-hydr}).

\section{Hydro+ near the QCD critical point}

\label{sec:hydro+-near-qcd}

\subsection{Formulation}
\label{sec:formulation}

In Section~\ref{sec:fluctuation-modes} we  identified slow degree(s) of freedom near the QCD critical point $\phi_\vQ$ and derived partial-equilibrium entropy density $s_{\plus}(\ed,n, \phi)$:
\begin{eqnarray}
\label{s+phi}
s_{\plus} (\ed, n, \phi_{\bm Q})
&=& s\(\ed, n\) + \frac{1}{2}\, \int_{\vQ}\, \left\{
 \log\frac{\phi_\vQ}{\bar\phi_\vQ(\ed,n)}
 -\frac{\phi_\vQ}{\bar\phi_{\vQ}(\ed,n)}
 +1\right\}\, ,
\end{eqnarray}
where $s(\e,n)$ is the ordinary equilibrium entropy and
$\bar\phi_\vQ(\ed,n)$ is the local equilibrium value of the
non-hydrodynamics slow mode $\phi_\vQ$ determined by local values of
$\ed$ and~$n$.  Similarly to $\ed$ and $n$ being energy and
charge density in the local rest frame defined by 4-velocity~$u^\mu$,
$\vQ$ is also a wave vector in the same frame, thus ensuring Lorentz
invariance of $s_\plus$. 
We are now ready to write the Hydro+ equations which we propose to describe the evolution near the critical point. 
The equations are usual conservation laws $\pd_{\mu}T^{\mu\nu}=0$, $\pd_{\mu}J^{\mu}=0$ and a relaxation rate equation~(\ref{eq:dtPhi}) for $\phi_\vQ$.
% { In principle, one could include the analogue of $A_{\phi}(Q)$ and $\l_{\a\pi}(Q), \l_{\pi\pi}(Q)$ term in Eq.~(\ref{eq:dtPhi}).
% We will defer this possibility to the future work.} 

The constitutive relation for $T^{\mu\nu}$ and $J^{\mu}$ now read:
\begin{eqnarray}
\label{Tmunu+}
  T^{\mu\nu} &=&  \ed u^\mu u^\nu + p_{(+)}\, g_\perp^{\mu\nu} 
   -\eta_{\plus} \(\pd_\perp^\mu u^{\nu} + \pd_\perp^\nu u^{\mu} 
- \frac23 g^{\mu\nu}_{\perp} \theta\) 
-\zeta_{\plus} g^{\mu\nu}_{\perp}\, \theta\, ,
\\
\label{Jmu+}
J^{\mu}&=& n\, u^{\mu} + \l\, \pd^{\mu}_{\perp}\, \a_{\plus}\, . 
\end{eqnarray}
Here $p_{\plus}$ is related to $s_{\plus}$ by
\begin{eqnarray}
\label{p+phi}
\b_{\plus} p_{\plus}= \, s_{\plus}-\b_{\plus} \e  +\a_{\plus}\, n \, ,
\end{eqnarray}

In this subsection we shall discuss the necessary ingredients for Hydro+. 
Our central ingredient is the partial-equilibrium equation of state,
which depends on the equilibrium entropy $s(\e,n)$ and the equilibrium value of $\bar\phi_\vQ(\ed,n)$. 
Here, $s(\ed,n)$ is the complete-equilibrium  equation of
state which includes the thermodynamic behavior near the critical point. 
For QCD, the equation of state in the relevant region, i.e., at finite
baryon density, is not reliably known from the first-principle lattice
simulation. An approach which is being pursued is to use an efficient
paramaterization of equation of state which, on the one hand matches
the reliable lattice data at small density (chemical potential) and on
the other hand incorporates correct universal critical behavior, in
order to minimize the number of free parameters to be determined by
comparing the simulation with experimental data~\cite{parotto}.

In this work we shall take $\phi_{\bm Q}$ to be the Wigner transform of the
correlator $\langle \delta m(\bm x) \delta m(\bm y) \rangle$. We shall assume
the separation of scales $1/\ell\ll Q \sim \xi^{-1}$, where $\ell$ is the
scale of spatial variation of local values of $\ed$, $n$, $u^\mu$ as well as
$\phi_{\bm Q}$.  The local equilibrium value, $\bar\phi_{\bm Q}$ is
given by the Wigner transform of the equilibrium correlator $\langle
\delta m(\bm x) \delta m(\bm 0) \rangle$ at given $\ed$ and $n$. In
the scaling regime ($\xi$ much larger than microscopic scale, such as $1/T$) the dependence of this quantity on $Q$ and $\xi$
must enter through a universal scaling function, i.e.,
\begin{equation}
  \label{eq:f2}
\bar\phi_{\bm Q}  = \int_{\Delta\bm x} e^{i{\bm  Q\cdot\bm x}}
\langle \delta m(\Delta\bm x) \delta m(\bm 0)\rangle 
= \bar\phi_{\bm 0} f_2(Q\xi,\Theta)
\end{equation}
where  $f_2$ is a universal function of two scaling variables with
$\Theta=\Theta(\ed,n)$ denoting the variable which, similarly to $\xi$, depends on
the local values of $\ed$ and $n$, but in contrast to $\xi$ is
invariant under scaling (e.g., parameter $\theta$ in $R\theta$
parameterization of the universal scaling equation of state \cite{Schofield:1969zza}).

The value of $\bar\phi_{\bm 0} $ is given by the magnitude of the
fluctuation of $m=\int_{\bm x} m(\bm x)/V$. This can be found from the standard text-book analysis of
the equilibrium fluctuations and is given by
\begin{equation}
  \label{eq:Deltam}
  \bar \phi_{\bm 0} = V\langle(\delta m)^2\rangle = \frac{c_p}{n^2}
\end{equation}
(see Appendix~\ref{sec:fluctuations-m-p} for derivation).

The universal function $f_2$ is normalized as $f_2(0,\Theta)=1$ and
can be determined, in principle, using methods described in, e.g.,
Ref.\cite{Kosterlitz1977}, or by a lattice computation. Similarly to
the equation of state $s(\ed,n)$ it is a {\em static\/} thermodynamic quantity.
The asymptotic behavior of
$f(x,\Theta)$ at large and small $x$ is given by
\begin{align}
  \label{eq:f2-asymp}
  &f_2 \to 1 + a_1 x^2 + \ldots, \quad x\ll1,\\
  &f_2 \to x^{-2+\eta}(b_1+b_2x^{-(1-\alpha)/\nu} +
    b_3x^{-1/\nu}\ldots), \quad x\gg1,
\end{align}
where coefficients $a_i$, $b_i$ are functions of
$\Theta$ (see Ref.~\cite{KosterlitzPRB}). 
%\footnote{The coefficient $b_1$ has to be a constant  multiple of $\phi_{\bm0}^{-1}$.}

For practical purposes a
simplified expression independent of $\Theta$
\begin{equation}
  \label{eq:f2-OZ}
  f_2(x) \approx (1+x^2)^{-1}
\end{equation}
often referred to as Ornstein-Zernike (OZ) form
\cite{KAWASAKI19701,RevModPhys.49.435} could provide a reasonable
approximation in limited applications. However, even though it
gives correct small-$x$ asymptotics, the incorrect large $x$
asymptotics leads to incorrect large-$\omega$ behavior of
frequency-dependent bulk viscosity which we discuss in
Section~\ref{sec:spectrum-kinetic}.

The coefficient $\gamma_\pi(\vQ)$ in Eq.~(\ref{eq:dtPhi}) is given, similarly to
Eq.~(\ref{G-phi-def}), by
\begin{equation}
  \label{eq:gpiQ}
  \gamma_\pi(\vQ) = \phi_\pi(\vQ) \Gamma(\vQ) 
= 2\bar\phi_\vQ^2 \Gamma(\vQ)\,,
\end{equation}
where we defined $\phi_\pi(\vQ)=(\pd\phi_\vQ/\pd\pi_\vQ)_{\ed n}$ and used
Eq.~(\ref{eq:piQ}).

The momentum-dependent relaxation rate of the critical slow mode
$\G(\vQ)$ for model H has been computed by Kawasaki
~\cite{KAWASAKI19701} (see also %Eq.(6.1.21) and 
Appendix 6B in
Ref.\cite{onuki2002phase} or Appendix~B of Ref.~\cite{Natsuume:2010bs} for derivations):
\begin{equation}
\label{kawasaki}
\G(\vQ) = 2\G_{\xi}\, K(Q\xi)\, ,
\end{equation}
where
\begin{equation}
K(x)=\(3/4\)\[ 1+x^{2}+\(x^{3}-x^{-1}\)\, \arctan(x)\]
\end{equation}
is sometimes referred to as Kawasaki
function.\footnote{\label{fn:factor2} The factor of
  2 relative to Ref.\cite{onuki2002phase} is due to the fact that
  $\Gamma(\vQ)$ is the relaxation rate of a 2-point function.} This
form describes the experimental data for various fluids over a range
of temperatures near the critical point remarkably well (see for
example Fig.~4 of Ref.~\cite{RevModPhys.49.435}).\footnote{The
  calculation of the Kawasaki function involves function $f_2$, and it
  is an example when the Ornstein-Zernike ansatz (\ref{eq:f2-OZ}) is
  adequate. } The characteristic critical slowing down rate is defined
as the diffusion rate at wave number $\xi^{-1}$
\begin{eqnarray}\label{eq:GD}
\G_\xi = D_p\, \xi^{-2}=\frac{T}{6\pi\eta\xi^3} 
\qquad
(z\approx 3)\, ,
\end{eqnarray}
where  $D_p$  is thermal (more precisely, entropy per baryon at
constant pressure)
diffusion constant due to the convection of critical density
fluctuations, the physical mechanism of which is
described in, e.g., Ref.\cite{RevModPhys.49.435}.

Finally, let us discuss the input values for Hydro+ kinetic
coefficients $\zeta_\plus$, $\eta_\plus$ and $\lambda$.
Since the critical behavior of bulk viscosity $\zeta$ (at zero
frequency) is now due to the dynamics of additional
slow mode $\phi_\vQ$, the input value $\zeta_{\plus}$ will not
sensitively depend on the correlation length and should match with the
smooth behavior of bulk viscosity away from the critical point. 
Since shear viscosity $\eta$ has very weak divergence at the critical
point, it is reasonable to neglect this divergence and 
use $\eta_\plus$ interpolated from the smooth behavior away from the critical point.

The implementation of conductivity $\l$ is more subtle.  $\l$ diverges
as $\xi$ (rounding the exponent to integer) due to the convection of
enhanced density
fluctuations~\cite{KAWASAKI19701,RevModPhys.49.435}. Somewhat similar to
divergence of $\zeta$, the divergence of $\lambda$ is due to slowness
of relaxation of a non-hydrodynamic mode involving transverse momentum and
charge density fluctuations. In principle,
this effect could be described by an extension of our approach using
2PI formalism by introducing an additional mode. 
We defer this to future work. As a provisional recipe, 
consistent with other choices, one could use, near the critical point,
\begin{equation}
  \label{eq:lambda}
  \lambda = \(\frac{nT}{w}\)^2c_p D_p
=  \(\frac{nT}{w}\)^2\frac{Tc_p}{6\pi\eta\xi}\,.
\end{equation}

\subsection{Frequency dependence of bulk response}
\label{sec:spectrum-kinetic}

In order to illustrate how the Hydro+ formalism described in
Section~\ref{sec:formulation} works, we shall study the bulk response in this
theory. It is useful to keep in mind that the formalism in
Section~\ref{sec:formulation} is essentially a multi-mode
generalization of the Hydro+ theory we discussed in
Section~\ref{sec:hydro-slow} and the results here are easily obtained by
generalizing calculation in Section~\ref{sec:plus-dispersion}.  The
purpose of this section is to demonstrate that this simpler formalism
reproduces known properties derived earlier using a different
approach by Kawasaki and Onuki~\cite{KAWASAKI19701,PhysRevE.55.403},
which we briefly review in the next section.

As in Sec.~\ref{sec:plus-dispersion}, we consider
expansion/compression fluctuations around the static and homogeneous
background.  A straightforward generalization of Eq. (\ref{eq:GR})
leads to the expression for the bulk response function
\begin{eqnarray}
\label{rho-kinetic0}
%\Delta\zeta(\omega) 
\frac{i\Delta G_R(\omega)}{\omega}
=\b\,\int_{\vQ}\, \frac{p^{2}_{\pi}(\vQ)}{\phi_{\pi}(\vQ)}\, 
\frac{1}{\G(\vQ)-i\o}\, ,
\end{eqnarray}
where we generalized the definition of $p_{\pi}, \phi_{\pi}$ for one
slow mode Eqs.~(\ref{eq:p-pi}),~(\ref{phi-pi-def}) to a branch of modes labeled by $\vQ$:
\begin{eqnarray}\label{eq:p-pi-Q}
p_{\pi}(\vQ) 
&\equiv& \(\frac{\delta p_{(+)}}{\delta\pi_{\bm Q}}\)_{\ed,n}
= - \frac{w}{\b}\, \(\frac{\partial\bar\phi_{\bm Q}}{\pd \ed}\)_{m}\, , 
\end{eqnarray}
\begin{eqnarray}\label{eq:phi-pi-Q}
\phi_{\pi}(\vQ) 
&=& \td{\phi_{\bm Q}}{\pi_{\bm Q}}_{\ed,n}
= 2\, \bar \phi_{\bm Q}^{2}\, .
\end{eqnarray}
In Eq.~(\ref{eq:p-pi-Q}) we have also used a natural generalization of Eq.~(\ref{p-pi}).
Consequently,
\begin{equation}
\label{rho-kinetic}
\frac{i\Delta G_R(\omega)}{\omega} =
\frac{w^{2}}{2\b}\,\int_{\bm Q}\,\(\frac{\pd\log\bar\phi_\vQ}{\pd \ed}\)^{2}_{m}\, \frac{1}{\G(\vQ)-i\o}\, . 
\end{equation}

Substituting the expression for the equilibrium mode distribution
$\bar\phi_\vQ$ we can determine the contribution of critical
fluctuations to frequency-dependent to bulk viscosity and the frequency
dependent equation of state stiffness (sound group velocity) which are
related to~$G_R$ as
\begin{equation}
  \label{eq:zetaGRcs}
  \Delta\zeta(\omega) = -{\rm Im}\,\D G_R/\omega;\quad
\Delta c_s^2(\omega) = {\rm Re}\, \D G_R/w.
\end{equation}
 Using the OZ
form in Eq.~(\ref{eq:f2-OZ}) one obtains the results in agreement with 
Kawasaki Ref.\cite{PhysRevA.1.1750}, which do not have the correct large
$\omega$ behavior. If one uses an ansatz satisfying the asymptotic
behavior in Eqs.~(\ref{eq:f2-asymp}) the correct asymptotic behavior
is reproduced, as in Ref.~\cite{onuki2002phase}.

To illustrate the frequency dependence of the bulk viscosity and
stiffness (sound speed)  we have chosen the ansatz for $\bar\phi_\vQ$ similar to the one
used by Onuki in Ref.~\cite{PhysRevE.55.403,onuki2002phase}. 
Rather than choosing the
function $\bar\phi_\vQ$, we chose ansatz for its derivative
$(\partial\log\bar\phi_\vQ/\partial\ed)^2_m\sim
x^2/(1+x^2)^{(1-\alpha)/\nu}$, where $x=Q\xi$, which satisfies the necessary
asymptotics following from Eqs.~(\ref{eq:f2-asymp}). In particular,
$(\partial\log\bar\phi_\vQ/\partial\ed)_m\sim Q^{-(1-\alpha)/\nu}$,
which translates into the large $\omega$ asymptotics
$G_R\sim\omega^{\alpha/(z\nu)}$ in accordance with Ref.\cite{onuki2002phase}.

\begin{figure}
\subfigure[]
{
\includegraphics[width=.45\textwidth]{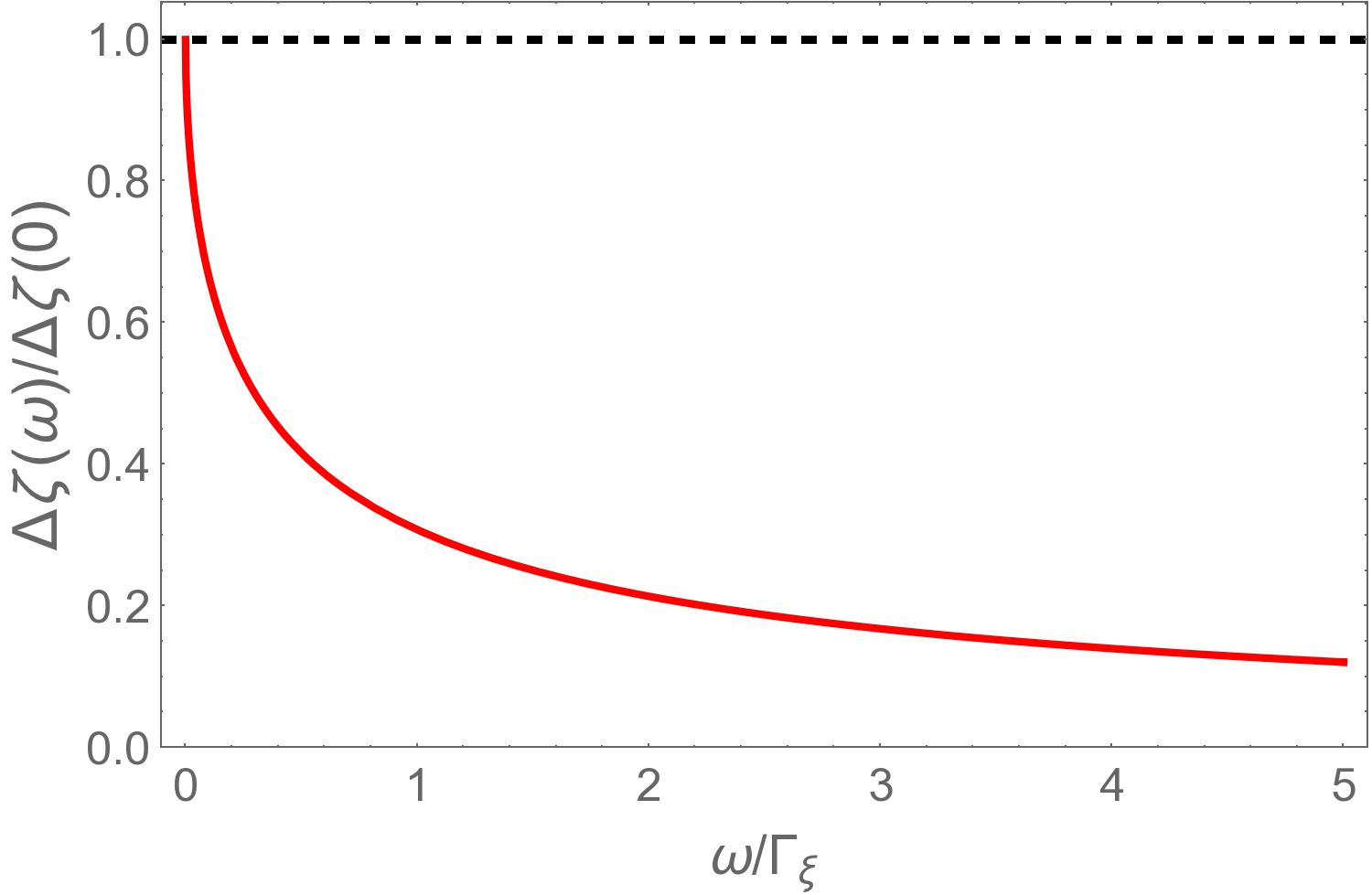}
\label{fig:ImGKin}
}
\subfigure[]
{
\includegraphics[width=.45\textwidth]{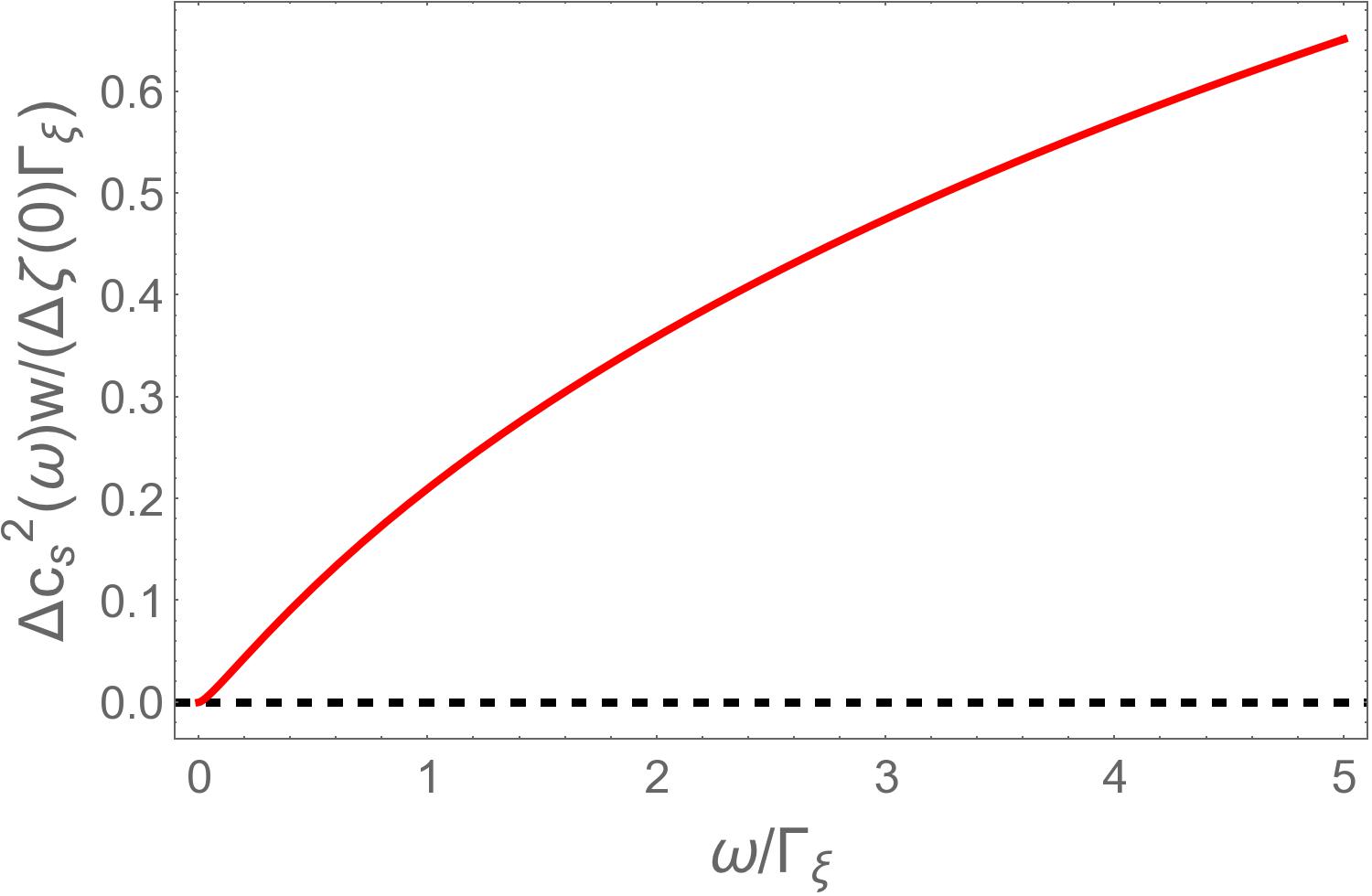}
\label{fig:ReGKin}
}
\caption{
\label{fig:PlotsGKin}
(color online) The frequency dependence of the (critical contribution
to) bulk viscosity, or $\Delta\zeta=-{\rm Im}\,G_R/\omega$ (left); and
the speed of sound (stiffness of equation of state), or
$\Delta c_s^2 = {\rm Re}\,G_R/w$, (right) near the critical point from
Eq.~(\ref{rho-kinetic}). The quantities and the frequency are
normalized to make plots scale-independent. Dashed horizontal lines
illustrate the results from ordinary hydrodynamics extrapolated beyond
its range of validity. }
\end{figure}

In Fig.~\ref{fig:PlotsGKin} we show the resulting frequency dependence
for $\Delta\zeta$ and $\Delta c_s^2$. Note that, as in
Section~\ref{sec:plus-dispersion}, the ordinary hydrodynamics
extrapolated beyond its region of validity overpredicts bulk viscosity
and underpredicts the stiffness.~\footnote{The stiffening at higher
  frequencies is a counterpart of the well-known effect of softening
  of the equation of state at $\omega\to0$, characterized by vanishing
  of the hydrodynamic sound speed $c_s^2$. More quantitatively, since
  $c_s^2\sim\xi^{-\alpha/\nu}$ the dynamical scaling translates this
  behavior into high frequency scaling $\omega^{\alpha/(\nu z)}$.}

\begin{figure}
\subfigure[]
{
\includegraphics[width=.45\textwidth]{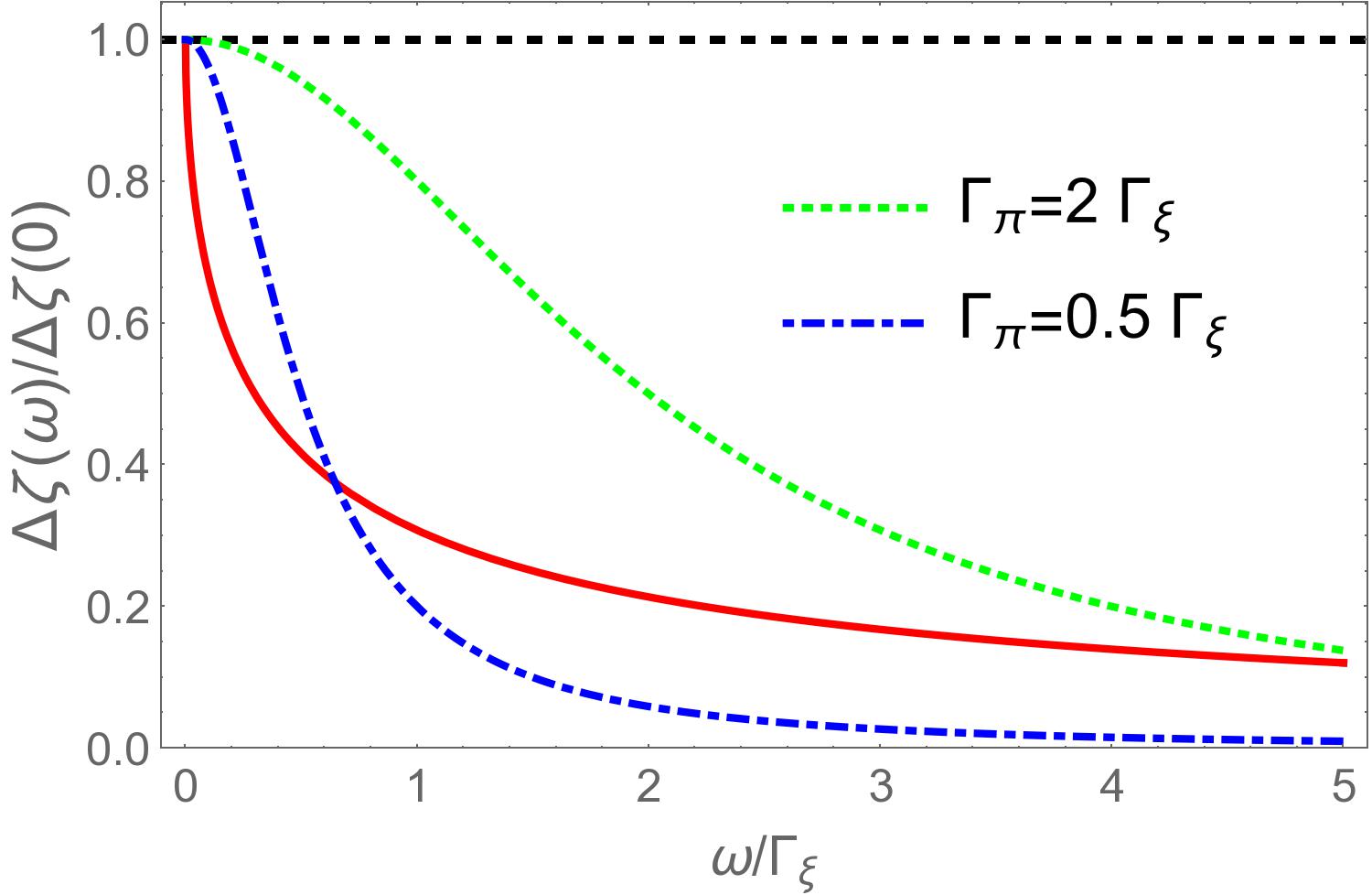}
\label{fig:ImCompare}
}
\subfigure[]
{
\includegraphics[width=.45\textwidth]{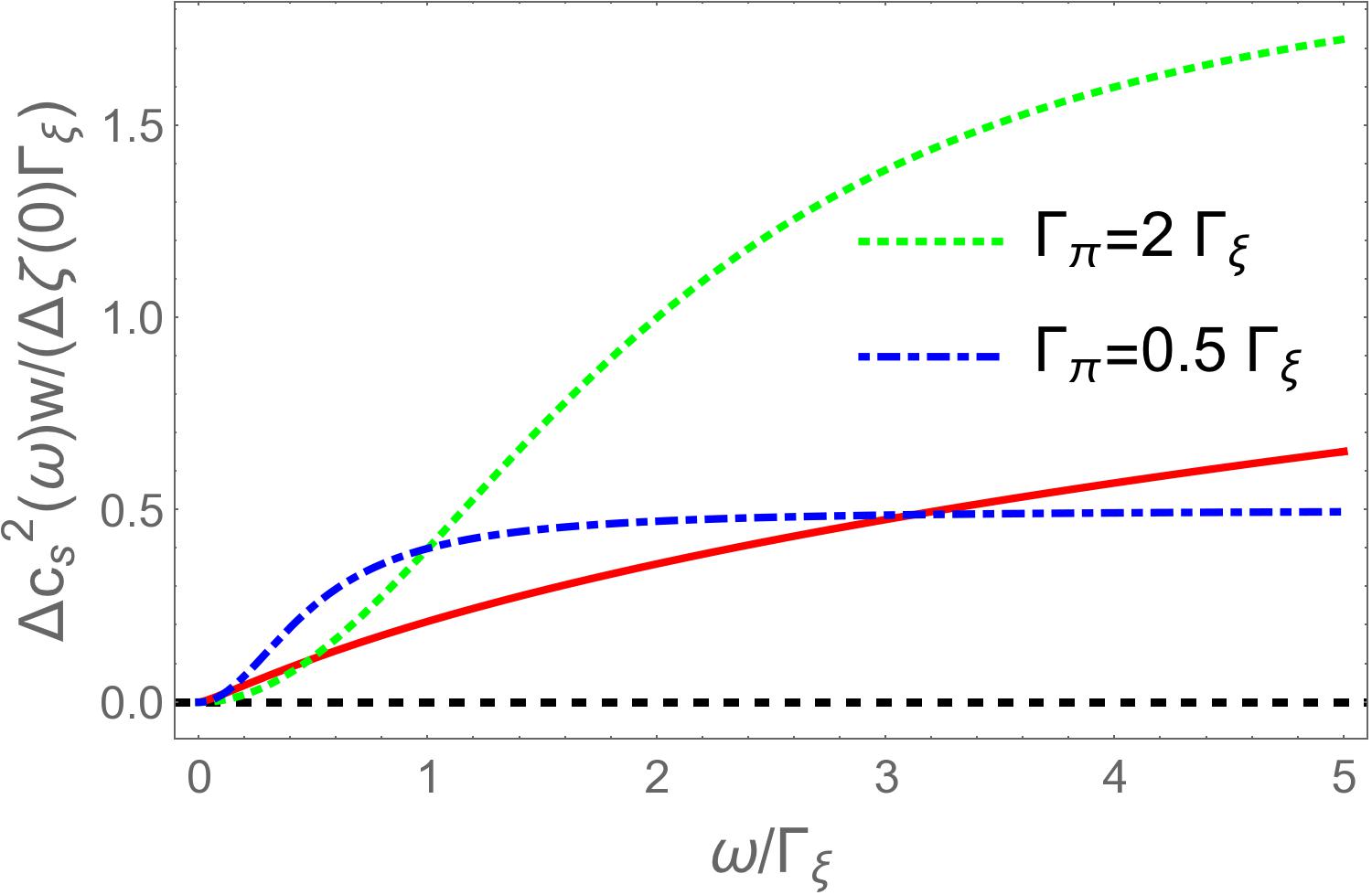}
\label{fig:ReCompare}
}
\caption{
\label{fig:PlotsCompare}
(color online) Comparison of the frequency dependence of the bulk viscosity, or
$\Delta\zeta$ (left); and the speed of sound (stiffness of
equation of state), or $\Delta c_s^2$, (right) between the two
theories: ``kinetic'' Hydro+ in Section~\ref{sec:formulation} and
single-mode Hydro+ in Section~\ref{sec:hydro-slow} for two choices of
the matching scale ratio $\Gamma_\pi/\Gamma_\xi$.
Dashed horizontal lines illustrate
the results of ordinary hydrodynamics extrapolated beyond its range
of validity. }
\end{figure}

Although these results are similar to the single-mode Hydro+ theory in
Section~\ref{sec:plus-dispersion}, 
it is instructive to compare and emphasize the differences. For this purpose we combine the plots from
Figs.~\ref{fig:GR1mode} and Figs.~\ref{fig:PlotsGKin} on the same
graph. To make the comparison we need to choose what scales in two
theories to match. The relevant scale in the single-mode theory is the
rate $\Gamma_\pi$ of the relaxation of the single mode $\phi$. In the
``kinetic'' Hydro+ theory there is a spectrum of modes, with a
characteristic scale given by $\Gamma_\xi$. We show two choices of
$\Gamma_\pi$-to-$\Gamma_\xi$ matching corresponding to
$\Gamma_\pi/\Gamma_\xi=0.5$ and $2$ in
Fig.~\ref{fig:PlotsCompare}. One can view this comparison as an answer
to the question: How well could a single-mode theory match the results
of a full ``kinetic'' Hydro+ approach. 

One can see that it is hard to
match both large and small frequency behavior of the bulk viscosity,
due to completely different asymptotics of $\Delta\zeta(\omega)$ in 
two theories.\footnote{The small-$\omega$ behavior $\Delta\zeta\sim -
  \omega^{-1/2}$ evident in Fig.~\ref{fig:ImGKin} is the half-integer
  long-time hydrodynamic
  tail~\cite{andreev1970twoliquid,Tail-Phys-Report,Kovtun:2003vj}. It cannot be matched by a
single-mode theory where $\Delta\zeta$ is analytic at zero frequency.} The same is true for the stiffness, i.e., $\Delta
c_s^2$. It is also notable that the choice of $\Gamma_\pi/\Gamma_\xi$
which makes $\Delta c_s^2$ agree better, could lead to worse agreement for
$\Delta \zeta$. These differences notwithstanding, it is also clear that
even a single-mode Hydro+ theory gives better description than naive
extrapolation of the ordinary hydrodynamics beyond its range of validity. 
 We can conclude that a single-mode theory could be used as a rough
illustration of some features of the critical slowing down, but it
cannot describe this phenomenon fully.

\subsection{Comparison to a loop calculation}
\label{sec:comp-loop-calc}

In the previous section, we have evaluated critical mode contribution
to the retarded Green's function
using Hydro+ formalism. To elucidate the correspondence between this
formalism and an earlier calculation by
Onuki~\cite{PhysRevE.55.403,onuki2002phase}, and to make the paper
self-contained, we shall provide here a sketch of the computation of
the same quantity using the approach of
Refs.~\cite{PhysRevE.55.403,onuki2002phase} 
(see also Ref.~\cite{Martinez:2017jjf,Akamatsu:2017rdu}) in this
subsection.

\begin{figure}
  \centering
  \includegraphics[width=16em]{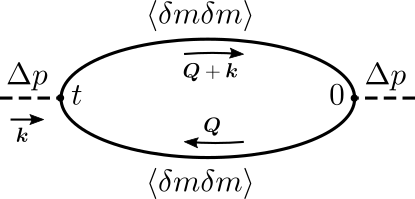}
  \caption{The one-loop diagram representing critical mode contribution to
    bulk response in Eq.~(\ref{eq:DG-dm4}). 
    Since $k\sim 1/\ell$, and $Q\sim1/\xi$, the separation of scales $\ell\gg\xi$ means $Q\gg
    k$ --- a typical hierarchy of scales in an HTL calculation.}
  \label{fig:loop}
\end{figure}

The starting point of Ref.~\cite{PhysRevE.55.403} is the relation between
$\Delta G_{R}(\o)$ and non-linear non-equilibrium pressure  $\D p$:
\begin{eqnarray}\label{eq:DG-DP}
%\zeta(\o)&=&
\frac{i\Delta G_{R}(\o)}{\o}= 
\b\, \int_0^\infty dt\, e^{i\o\,t}\,\int_{\vx}\,
 \<\, \D p\(t,\vx\)\, \D p(0,\bm0)\,\>\, ,
\end{eqnarray}
where we used the fluctuation-dissipation relation. The
non-linear non-equilibrium pressure can be related to corresponding
contribution to entropy density, $\Delta s$:
\begin{eqnarray}\label{eq:Dp-Ds}
\b \, \D p = -w\, \D\beta + n \D \a
= - w\(\frac{\pd \D s}{\pd \ed}\)_{\!m}\, ,
\end{eqnarray}
where we used
\begin{equation}
\D \b = \(\frac{\pd \D s}{\pd \ed}\)_{\!n}\, ,
\qquad
\D \a =- \(\frac{\pd \D s}{\pd n}\)_{\!\ed}\, 
\end{equation}
and thermodynamic relation (\ref{eq:ed/n=w/n}).
Near the critical point, it is sufficient to keep only the
contribution of the slowest mode $\d m$ to $\D s$:
\begin{eqnarray}\label{eq:Ds-dmdm}
\int_{\vx} \D s = -\frac{1}{2}\int_\vQ \, \bar{\phi}_\vQ^{-1}\, |\d m_\vQ(t)|^{2}\, . 
\end{eqnarray}
where $\bar\phi_\vQ$ is the equal-time correlator of $\delta m$ as in
Eq.~(\ref{eq:f2}) and $\delta m_\vQ$ is the Fourier transform of $\delta
m(\vx)$. Substituting Eqs.~(\ref{eq:Dp-Ds})
and~(\ref{eq:Ds-dmdm}) into Eq.~(\ref{eq:DG-DP}) we find (using the
scale separation $1/\ell\ll Q$)
\begin{equation}\label{eq:DG-dm4}
%\zeta(\o)&=&
\frac{i\Delta G_{R}(\o)}{\o}=
\frac{w^2}{2\b}\, \int_0^\infty dt\, e^{i\o\,t}%\,\int\! d\vx\,
% \<\, \D p\(t,\vx\)\, \D p(0,\bm0)\,\>\, ,
\(\frac{\bar{\phi}_\vQ^{-1}}{\pd \ed}\)^{2}_{\!m}
\<  \delta m_\vQ(t)\, \delta m_{-\vQ}(0)\>\,
\<  \delta m_{-\vQ}(t)\, \delta m_\vQ(0)\>
\end{equation}
This result is best illustrated by a simple one-loop diagram shown in
Fig.~\ref{fig:loop}. The vertex factor,
$(\pd\bar\phi_\vQ^{-1}/\pd\ed)$, is a third derivative of the entropy
$\Delta s$ (see Eq.~(\ref{eq:Ds-dmdm})), which is intuitively natural
considering analogy between the entropy of fluctuations and the action
in field theory.

Using the expression for the unequal-time correlator 
\begin{equation}
  \label{eq:dmdmt}
  \<\,\delta m_\vQ(t)\, \delta m_{-\vQ}(0) \,  \> = \bar\phi_\vQ\, e^{-\Gamma(\vQ)t/2}
\end{equation}
and performing the time integration in Eq.~(\ref{eq:DG-dm4}) we obtain
precisely Eq.~(\ref{rho-kinetic}). The factor $1/2$ in the exponent in
Eq.~(\ref{eq:dmdmt}) is due to the fact that $\Gamma(\vQ)$ is the
relaxation rate of a {\em two}-point function (essentially, of
$(\delta m)^2$), while Eq.~(\ref{eq:dmdmt}) represents the
correlation function of the one-point function $\delta m$ (see also
Eq.~(\ref{kawasaki}) and footnote~\ref{fn:factor2}).

The comparison between Hydro+ and the loop calculation in this Section
sheds additional light on the analogy between Hydro+ and the kinetic
(Boltzmann) description in thermal field theory. The response in
thermal field theory, which can be calculated in the hard-thermal loop
(HTL) approach, is non-local, as manifested by the well-known Landau
damping. This response is due to almost on-shell propagation of weakly
coupled (quasi)particles. However, one can replace HTL approach by an
equivalent kinetic description of these particles using a Boltzmann
equation (or, in the case of gauge-field dynamics, Vlasov equations,
coupling particles to classical fields). The advantage of the kinetic
approach is that it is local and thus more intuitive and conceptually
satisfying. Locality is also an indispensable property for numerical
simulations of real-time dynamics. Hydro+ is similar to kinetic theory
in this respect. While non-instantaneous bulk response (e.g.,
frequency-dependent bulk viscosity) is hard to implement in a
simulation directly, Hydro+ reproduces this phenomenon using local
(instantaneous) dynamics of additional modes very similar to kinetic
description.

\section{Summary and discussion}

We considered an extended hydrodynamic theory, or Hydro+, which
describes evolution of partial-equilibrium states characterized by
off-equilibrium values of non-hydrodynamic but slow variables.  In
general, such an extension can be justified if the additional
non-hydrodynamic variable is still much slower than the remaining
microscopic variables which are left (integrated) out. This condition
distinguishes our approach from other extended hydrodynamic proposals,
such as, e.g., the well-known Israel-Stewart second-order
hydrodynamics~\cite{Israel:1979wp}, where additional variables
(components of stress tensor), in general, relax to equilibrium as
fast as the other (infinitely many) microscopic modes. As a result,
applicability of Israel-Stewart theory (beyond ordinary hydrodynamic
regime) is questionable~\cite{Geroch:2001xs}. Unlike Israel-Stuart
hydrodynamics, we wish to consider a systematic limit in which the
variables we keep are slow either because they are conserved (i.e.,
hydrodynamic), or because there exists another parameter, independent
of the scale of inhomogeneity $\ell$, controlling the slowness of
hydrodynamic
variables.

In Section~\ref{sec:hydro-slow} we describe a general formulation of
Hydro+ and discuss the chiral (anomalous) fluid as a simple example in
Appendix~\ref{sec:nA}. We show how a competition between scales of
hydrodynamic evolution and non-hydrodynamic slow mode relaxation
gives rise to two distinct regimes of frequencies. For
$\omega\ll\Gamma_\pi$ the slow mode it completely equilibrated and
simply tracks the hydrodynamic variables. Ordinary hydrodynamics apply
in this regime, but the effect of the slow mode is manifested in a large
contribution to bulk viscosity proportional to $1/\Gamma_\pi$ -- a
phenomenon already known in the context of non-relativistic fluids~\cite{mandelshtam1937theory,landau1954anomalous,landau2013fluid}.
For $\omega\gg\Gamma_\pi$ the slow mode is effectively ``frozen'', which
leads to a different, stiffer equation of state and the drop of the bulk
viscosity. 

The phenomenon of critical slowing down near a critical point is very
similar to the situation where Hydro+ is applicable. The slowest
non-hydrodynamic relaxation rate is controlled by the value of the
correlation length $\xi$ i.e., $\Gamma_\xi\sim\xi^{-3}$, independently
of the scale of inhomogeneity $\ell$ which controls the slowness of
the conserved hydrodynamic modes, $\Gamma_{\rm hydro}\sim\ell^{-2}$
for relaxation or $\ell^{-1}$ for propagation. This sets the stage for
the scale competition characteristic of Hydro+.

Ordinary hydrodynamics breaks down when
$\Gamma_{\rm hydro}\gtrsim \Gamma_\xi$, since the condition of the
separation of hydrodynamic and non-hydrodynamic relaxation rates is
violated leading to non-locality of the theory. Our goal in this paper
is to determine how to add the slow mode, or modes, to hydrodynamics
in order to extend its validity to the regime where the evolution
rate of the hydrodynamic modes is comparable to, or larger than, the
relaxation rate $\Gamma_\xi$.

The central element of Hydro+ is the extended equation of state,
given by entropy, $s_\plus$, of the partial-equilibrium state
characterized by the values of the non-hydrodynamic mode as well as
the hydrodynamic variables. One of the major results of this paper is
the expression for the entropy as a function(al) of the
off-equilibrium values of {\em fluctuations}. In
Section~\ref{sec:entr-fluct-2pi} we derive a general expression which
we find to bear a natural mathematical resemblance to the 2PI action
in quantum field theory. In addition to one-point functions --
the local values of the hydrodynamic variables -- the 2PI entropy
depends on the 2-point functions -- the off-equilibrium values of the
fluctuations. In Section~\ref{sec:fluctuation-modes} we use the
extended entropy to write Hydro+ equations generalizing the
single-mode theory discussed in Section~\ref{sec:hydro-slow}.

In Section~\ref{sec:hydro+-near-qcd} we focus on the slowest mode near
the critical point -- diffusion of entropy per (baryon) charge, and
write down a set of equations for the coupled evolution of critically
slow magnitude of fluctuations and the hydrodynamic modes. Since the
additional slow variable is a measure of fluctuations, i.e., a 2-point
function, the corresponding variable has an index, $\bm Q$, which is
the Fourier transform of the spatial separation of the points in the
2-point function. In that respect the slow variable $\phi_\vQ$ is similar to the
phase-space particle distribution function in kinetic theory and can
be called mode distribution function.

We show that the Hydro+ theory defined in
Section~\ref{sec:formulation} reproduces known phenomena associated
with critical slowing down. In particular, we show that the bulk
viscosity receives critically enhanced contribution
$\Delta\zeta\sim \xi^3$. As a result, ordinary hydrodynamics breaks
down when the bulk relaxation (sound attenuation) rate
$\Gamma_{\rm hydro}$ is of order $\Gamma_\xi\sim\xi^{-3}$. Naively
extending hydrodynamics beyond that (to larger frequencies or wave
numbers) one overpredicts the actual amount of dissipation because the
frequency-dependent bulk viscosity drops for larger frequencies. This
phenomenon is captured by Hydro+ as illustrated by
Fig.~\ref{fig:ImGKin}.

We also note that as a consequence of the critical slowing down
the stiffness of the equation of state increases for frequencies above
$\Gamma_\xi$. Again, the ordinary hydrodynamics will underpredict the
stiffness (measured by the frequency dependent sound velocity) as
shown in Fig.~\ref{fig:ReGKin}.

The purpose of our paper is to introduce the approach and discuss its
advantages as well as to point out  limitations which could be
addressed in  future work.

One of the advantages of the approach to fluctuations encoded in
Hydro+ over another popular approach being discussed in the literature
based on the stochastic
hydrodynamics~\cite{Kapusta:2011gt,Young:2014pka} is that Hydro+
eliminates the need for solving stochastic equations. Even though two
approaches lead to similar results, deterministic description of
fluctuations may prove advantageous in numerical simulations. One of
the reasons is that stochastic fluctuations introduce strong cutoff
dependence of equation of state as well as kinetic coefficients as has
been observed and discussed, e.g., in
Refs.\cite{andreev1978corrections,Akamatsu:2017rdu}. This dependence need to be
canceled, which creates a numerically ill-conditioned problem. On the other
hand, the 2PI equation of state we introduce in Section~\ref{sec:entr-fluct-2pi} is already
renormalized (see Eq.~(\ref{eq:4})), i.e., the UV divergences due to
fluctuations are included into it.

This paper would be incomplete without a discussion of the domain of
applicability of Hydro+. As we already pointed out, ordinary
hydrodynamics breaks down for frequencies (or rates) larger than the
rate $\Gamma_\xi$ which becomes critically slow as
$\Gamma_\xi\sim\xi^{-3}$. Hydro+ extends the range of applicability to
higher frequencies. Unlike the simple single-mode Hydro+ discussed in
Section~\ref{sec:hydro-slow}, which could be applicable all the way to
the microscopic scale (collision rates, or $T$), Hydro+ near the
critical point has another limitation -- the rate of the relaxation of
the next-to-slowest mode. This rate is {\em parametrically\/} faster than
$\Gamma_\xi$, but still slower than microscopic scale, e.g., $T$. The
corresponding mode is the relaxation of the transverse velocity
(shear) fluctuations on the characteristic scale $\xi$ of
density fluctuations, with the rate
$\Gamma^T_\xi=(\eta/w)\xi^{-2}\sim\xi^{-2}$.\footnote{Here, for
  simplicity, we neglect $\xi$-dependence of $\eta$, a common
  approximation, since it is very weak \cite{RevModPhys.49.435}.}
Comparing $\Gamma^T_\xi$ to $\Gamma_\xi$ we see that Hydro+ extends
the range of applicability by a factor
$\Gamma^T_\xi/\Gamma_\xi\sim T \xi \gg 1$.

The emergence of the scale $\Gamma_{\xi}^T$ is due to the important
role played by the transverse modes in the Model H dynamic
universality class~\cite{RevModPhys.49.435}. Indeed, the
characteristic rate $\Gamma_\xi$ in Eq.~(\ref{eq:GD}) (and the
corresponding value of the critical exponent $z\approx 3$) depends on
the divergence of conductivity $\l\sim \xi$.  This critical behavior
of $\l$ is driven by the enhanced fluctuations of charge density and
relies on transverse velocity (shear) modes relaxing faster than
$\omega$. Therefore, $\l$ will only reach its critical behavior
$\l\sim \xi$ for processes much slower than $\Gamma^T_\xi$ and the use
of Eq.~(\ref{eq:GD}) (and $z\approx 3$) is only meaningful for
$\omega\ll\Gamma_\xi^T$.  To extend Hydro+ to time scales shorter than
$1/\Gamma_{\xi}^{T}$ on would need to add the fluctuations of shear
modes as additional non-hydrodynamic variables, which should be
possible to do
along the lines similar to Section~\ref{sec:fluctuation-modes}.

The 2PI formalism we introduce in Section~\ref{sec:entr-fluct-2pi} is
suited for treating the off-equilibrium evolution of Gaussian
fluctuations. The sensitivity of {\em non}-Gaussian measures of
fluctuations makes them important signatures of the QCD critical
point~\cite{Stephanov:2008qz,Stephanov:2011pb}. In order to
incorporate the evolution of non-Gaussianity into Hydro+ formalism one
needs to extend 2PI formalism to non-Gaussian fluctuations. This would
lead to a generalization of 2PI entropy to, e.g., 3PI and
4PI, and a hierarchy of kinetic equations similar to the hierarchy of
cumulant equations in Ref.~\cite{Mukherjee:2015swa}. We defer these
and other developments of Hydro+ to future work.

\appendix

\section{Application of Hydro+ to chiral (anomalous) fluid
\label{sec:nA}
}

We consider a chiral fluid, the constituents of which include (approximately) massless fermions. 
In this system, 
the axial current $J^{\mu}_{A} \equiv
\bar{\psi}\g^{\mu}\g^{5}\psi$ is conserved only approximately. The
conservation is violated by a small fermion mass (and/or by quantum
anomaly, as 
in non-Abelian gauge theories, 
where topological sphaleron fluctuations induce fermion chirality flips).
Therefore:
\begin{equation}
\label{dJA}
\pd_{\mu}\, J^{\mu}_{A} =- \g_{A}\, \a_{A}\, .
\end{equation}
Here $\g_{A}$ is an Onsager coefficient describing chirality
violating processes and $\a_{A}=\b\mu_{A}$ where $\mu_{A}$ is the axial chemical potential. 
In the absence of a background magnetic field $B^{\mu}$, the currents
$\D J^{\mu}_{V}$ and $\D J^{\mu}_{A}$ are given by:
\begin{equation}
\label{JVA}
\D J^{\mu}_{V}
= %C_{A}\, \frac{\a_{A}}{\b}\, B^{\mu} + 
\l_{VV}\,  \pd^{\perp}_{\mu}\a_{V} + \l_{VA}\,  \pd^{\perp}_{\mu}\a_{A}\, , 
\qquad
\D J^{\mu}_{A}
=% C_{A}\, \frac{\a_{V}}{\b}\, B^{\mu} +
 \l_{VA}\,  \pd^{\perp}_{\mu}\a_{V} + \l_{AA}\, \pd^{\perp}_{\mu}\a_{A}\, .
\end{equation}
%We have included currents due to anomalous transport. 
Parameters $\l_{VV}$, etc.\ are vector and axial conductivity coefficients
(conductivity times temperature). 
Substituting \eqref{JVA} into \eqref{dJA}, 
we then have:
\begin{eqnarray}
\label{DnV}
D\, n_{V}&=& - n_{V}\, \theta
 -\partial_\perp \cdot \(\,\lambda_{VV}\,\partial\alpha_{V} +\lambda_{VA}\,  \partial\a_{A}\,\)
\\
\label{DnA}
D\, n_{A}&=& - n_{A}\, \theta - \g_{A}\, \a_{A}
-\partial_\perp \cdot \(\,\lambda_{AA}\,\partial\alpha_{A} +
  \lambda_{V A}\,  \partial\a_{V}\,\)
\, . 
\end{eqnarray}
We now identify axial charge density $n_{A}$ with $\phi$.  
Comparing \eqref{DnA} with \eqref{phi-eq} and \eqref{eq:Fphi}, 
we have $A_\phi\equiv n_{A}$, and $ \g_{\pi}\equiv\g_{A}$. 
 In general, with  $n_{A}$ finite, $\l_{AV}$ is non-zero. 
Moreover, $\l_{AA}$ and $\l_{VV}$ can be different from each other. 
Such expectation has been confirmed in an explicit perturbative computation~\cite{Huang:2013iia}. 
 
In this theory the analog of $p_\pi$ is zero due to
parity and thus bulk viscosity and sound propagation velocity are not
affected by the slow mode as it is in a more general case described in
Section~\ref{sec:hydro-slow}.  However, a similar enhancement was found
in Ref.~\cite{Stephanov:2014dma} for the conductivity along the
magnetic field, which receives anomalous contribution from the slow
non-hydrodynamics mode:
\begin{eqnarray}
\Delta \lambda = \chi_V T\,\frac{\D v^{2}_{\rm cmw}}{\Gamma_{\pi}}\, ,
\end{eqnarray}
where $\chi_V$ is the charge susceptibility. Similarly to Eq.~(\ref{eq:LK}),  $\D v^{2}_{\rm cmw}$ denotes the increase of the speed of
chiral magnetic wave between hydrodynamic regime $\omega\ll\Gamma_\pi$
and Hydro+ regime $\omega\gg\Gamma_\pi$, and $\Gamma_\pi$ is the
rate of the slow mode (axial charge) relaxation.

\section{A thermodynamic relation for $p_\pi$}
\label{sec:therm-relat-p_pi}

This section supplies derivation of Eq.~(\ref{eq:p-pi-phi-ed}),
which can be also written as
\begin{eqnarray}
\label{p-pi}
 \(\frac{\pd p_{\plus}}{\pd \pi}\)_{\!\ed n} 
=- \frac{w}{\b}\, \[
  \(\frac{\pd\bar\phi}{\pd\ed}\)_{\!m}-\frac{A_{\phi}}{w}\]\, 
\quad\mbox{at $\pi=0$}
\end{eqnarray}
due to Eq.~(\ref{eq:n/s-en}), which means
\begin{equation}
  \label{eq:ed/n=w/n}
  \td{n}{\ed}_{\!m} = \frac{n}{w} \,.
\end{equation}

We begin from (\ref{beta-dp}) evaluated at $\pi=0$:
\begin{equation}
  \label{eq:beta-dp-0}
  \b\, d p_{\plus} = -w\, d\b_{\plus} + n\,
  d\a_{\plus}  + A_{\phi} d\pi \, . 
\end{equation}
Considering variation of $\pi$ at $\ed$ and $n$ fixed we can write
\begin{equation}
  \label{eq:ppi-betapi}
  \b\, p_\pi = -w \b_\pi + n \a_\pi + A_\phi
\end{equation}
where index $\pi$ denotes derivatives with respect to $\pi$ at  $\ed$
and $n$ fixed evaluated at $\pi=0$, e.g.,
\begin{equation}
  \label{eq:p-pi-def}
  p_\pi \equiv  
\(\frac{\pd p_{\plus}(\ed,n,\pi) }{\pd \pi}\)\Bigg |_{\pi=0}
\equiv
\(\frac{\pd p_{\plus} }{\pd \pi}\)_{\!\ed,n}\Bigg |_{\pi=0}\,.
\end{equation}

 We can then use Maxwell relations
applied to the differential
\begin{equation}
  \label{eq:s+piphi}
  d\(s_\plus + \pi\phi\) = \b_\plus d\ed - \a_\plus dn + \phi d\pi
\end{equation}
to relate derivative w.r.t. $\pi$ to the derivatives of $\phi$ (at
$\pi=0$, i.e., in equilibrium):
\begin{equation}
  \label{eq:phi/ed-phi/n}
  \b_\pi \equiv \td{\b_\plus}{\pi}_{\ed n} = \td{\phi}{\ed}_{\pi n},
\qquad
\a_\pi \equiv \td{\a_\plus}{\pi}_{\ed n} =  -\td{\phi}{n}_{\pi\ed}.
\end{equation}

Substituting Maxwell relations (\ref{eq:phi/ed-phi/n}) into
Eq.~(\ref{eq:ppi-betapi}) we obtain Eq.~(\ref{eq:p-pi-phi-ed}) or
(with Eq.~(\ref{eq:ed/n=w/n})) Eq.~(\ref{p-pi}).

\section{Reparameterization covariance in Hydro+}
\label{sec:repar-covar-hydr}

The choice of the slow mode is not unique, but since the physics
cannot depend on that choice the equations of Hydro+ must possess
reparameterization invariance which we describe here.

Let us consider another choice of the slow variable, $\phi'$, which is
a function of the original choice $\phi$ and, possibly, of $\ed$ and
$n$, i.e.,
\begin{equation}
  \label{eq:phi'}
  \phi' = f(\phi,\ed,n).
\end{equation}
The equation governing evolution of $\phi'$ must have a similar form
to Eq.~(\ref{phi-eq}), i.e.,
\begin{equation}
  \label{eq:Dphi'}
  D\phi' = - F_\phi' - A_\phi'\theta,
\end{equation}
where the relationship between the new parameters $F_\phi'$ and
$A_\phi'$ and the original ones can be found by substituting
Eq.~(\ref{eq:phi'}) into Eq.~(\ref{eq:Dphi'}) and matching to
Eq.~(\ref{phi-eq}):
\begin{equation}
  \label{eq:FA}
  F_\phi' = f_\phi F_\phi,\qquad A_\phi' = f_\phi A_\phi +
  w f_\ed + n f_n
\end{equation}
where
\begin{equation}
f_\phi = \td{f}{\phi}_{\ed n}
\quad f_\ed\equiv\td{f}{\ed}_{n\phi};
\quad f_n\equiv \td{f}{n}_{\ed\phi}\label{eq:f_enphi}
\end{equation}
One can easily check that
\begin{equation}
  \label{eq:pi'}
  \pi' = f_\phi \pi;
\quad \phi_\pi' = f_\phi^2\phi_\pi; 
\quad \gamma_\pi' = f_\phi^2\gamma_\pi;
\quad p_\pi' = f_\phi p_\pi.
\end{equation}
These transformations leave the relationship Eq.~(\ref{p-pi}) between
$p_\pi$ and $\partial \bar\phi/\partial\ed$ invariant (note that
transformation of $A_\phi$ and the role it plays is reminiscent of a
gauge potential). Also, the combination $p_\pi^2/\phi_\pi$ in the
definition of $\Delta c_s^2$ is invariant as it should be expected.

\section{Fluctuations of $m$ and $p$}
\label{sec:fluctuations-m-p}

The fluctuations of the thermodynamic quantities $\Psi$ are described
by the probability distribution $P\sim e^{S(\Psi)+\bar J\Psi}$ (see
Section~\ref{sec:entr-fluct-2pi} and Ref.~\cite{Lifshitz:v5}). The
change of $P$ under a fluctuation $\Psi-\bar \Psi\equiv\delta\Psi$ (of
arbitrary, not necessarily small, magnitude) is
given by
\begin{equation}
  \label{eq:P-fl}
 \Delta (\log P) \equiv \log \frac{P(\Psi)}{P(\bar\Psi)}
=  S(\bar\Psi+\delta\Psi) - S(\Psi) + \bar J\delta \Psi
 \equiv \Delta S + \bar J\delta \Psi\,,
\end{equation}
where $\Delta S$ is the change of the entropy of the system and $\bar
J\delta\Psi$ is the change of the entropy of the environment as the
amounts $\delta\Psi$ of
{\em conserved} quantities are exchanged between the system and
the environment characterized by thermodynamic potentials $\bar
J$. Since $\bar \Psi$ is the equilibrium value, the $\mathcal
O(\delta\Psi)$ terms in Eq.~(\ref{eq:P-fl}) cancel. For the case
we consider $\Psi=(\ed,n)$ and $J=(-\beta,\alpha)$, Taylor expanding
the entropy to second order in $\delta\Psi$ we can write
\begin{equation}
  \label{eq:P-fl2}
 \Delta (\log P)/V = \frac{1}{2}\left(
\delta\beta\delta\ed - \delta\alpha\delta n
\right) + \mathcal O (\delta^3)\,,
\end{equation}
where the factor of volume $V$ comes from the space integration.
Expressing the variables $p$ and $m=s/n$ in terms of $\beta$, $\alpha$, $\ed$, $n$
one finds to linear order:
\begin{align}
  \label{eq:dpdm}
  \beta\delta p &= n \delta \alpha - w \delta\beta + \mathcal O (\delta^2)\\
  n^2\delta m   &= \beta (n \delta \ed - w \delta n)+ \mathcal O (\delta^2)
\end{align}
Solving for $\delta\ed$ and $\delta \alpha$ and substituting into
Eq.~(\ref{eq:P-fl}) one finds (upon cancellation of $\delta\beta\delta
n $ terms):
\begin{equation}
  \label{eq:logP-dpdm}
\Delta (\log P)/V = \frac{\beta n}2  \left(
 \delta\left(\frac1n\right)\,\delta p
-
\delta\left(\frac1\beta\right) \delta m
\right) + \mathcal O(\delta^3)
\end{equation}
Expressing $\delta(1/n)$ and $\delta(1/\beta)$ in terms $\delta m$ and
$\delta p$ and using the Maxwell relations stemming from
\begin{equation}
  \label{eq:dw/n}
  d\left(\frac{w}{n}\right) = \frac1\beta\,dm + \frac1n\,dp
\end{equation}
one arrives at

\begin{equation}
  \label{eq:logP-dpdm-2}
\Delta (\log P)/V = 
- \frac12\left(
\frac{\beta}{wc_s^2}(\delta p)^2
+\frac{n^2}{c_p}(\delta  m)^2
\right)
+ \mathcal O(\delta^3)\,.
\end{equation}

In the thermodynamic limit, i.e., for large $V\sim k^{-3}\gg\xi^3$ (or
$\ell\gg\xi$), fluctuations are small, their probability
distribution is approximately Gaussian, and we obtain
Eq.~(\ref{eq:Deltam}) for their variance.

\begin{acknowledgments}
  We thank Tigran Kalaydzhyan for collaboration during the initial
  stage of this project.  We thank Boris Spivak for drawing our attention to Ref.~\cite{andreev1970twoliquid}. 
  We thank Ulirch Heinz, Krishna Rajagopal, Derek Teaney, Ho-Ung Yee for helpful discussions.
  This work is supported by the U.S. Department of Energy, Office of
  Science, Office of Nuclear Physics, within the framework of the Beam
  Energy Scan Theory (BEST) Topical Collaboration and grants
  Nos. DE-FG0201ER41195 (M.S.) and DE-SC0011090 (Y.Y).
\end{acknowledgments}

\bibliography{hydro+}

\end{document}